\documentclass[conference]{IEEEtran}

\usepackage{times}
\usepackage{epsfig}
\usepackage{graphicx}
\usepackage{amsmath}
\usepackage{amssymb}
\usepackage{multirow}
\usepackage{makecell}
\usepackage{subcaption}
\usepackage{cite}
\usepackage{cleveref}
\usepackage{graphicx}
\usepackage[font=small,labelfont={bf}]{caption}
\usepackage{placeins}
\usepackage[aboveskip=0pt,font=scriptsize]{subcaption}
\usepackage{xcolor}
\usepackage{enumitem}
\usepackage{float}
\usepackage{multirow}
\usepackage{makecell}
\usepackage{booktabs}
\newcommand\crule[3][black]{\textcolor{#1}{\rule{#2}{#3}}}
\usepackage{lipsum}

\IEEEoverridecommandlockouts

\DeclareRobustCommand*{\IEEEauthorrefmark}[1]{%
  \raisebox{0pt}[0pt][0pt]{\textsuperscript{\footnotesize #1}}%
}

\begin{document}

\title{An Explainable Adversarial Robustness Metric for Deep Learning Neural Networks}


\author{\IEEEauthorblockN{Chirag Agarwal${}^{*}$\IEEEauthorrefmark{1},
Bo Dong${}^{*}$\IEEEauthorrefmark{2},
Dan Schonfeld\IEEEauthorrefmark{1},
Anthony Hoogs\IEEEauthorrefmark{2}}\\
\IEEEauthorblockA{\IEEEauthorrefmark{1}Department of Electrical and Computer Engineering, 
University of Illinois at Chicago, Illinois.\\ Email: [cagarw2, dans]@uic.edu}
\IEEEauthorblockA{\IEEEauthorrefmark{2}Kitware Inc.,Clifton Park, New York.\\
Email: [bo.dong, anthony.hoogs]@kitware.com}
\thanks{\textbf{${}^{*}$Authors colloborated equally and are ordered by their last names.}}}

\maketitle

\begin{abstract}
Deep Neural Networks(DNN) have excessively advanced the field of computer vision by achieving state of the art performance in various vision tasks. These results are not limited to the field of vision but can also be seen in speech recognition and machine translation tasks. Recently, DNNs are found to poorly fail when tested with samples that are crafted by making imperceptible changes to the original input images. This causes a gap between the validation and adversarial performance of a DNN. An effective and generalizable robustness metric for evaluating the performance of DNN on these adversarial inputs is still missing from the literature. In this paper, we propose Noise Sensitivity Score (NSS), a metric that quantifies the performance of a DNN on a specific input under different forms of fix-directional attacks. An insightful mathematical explanation is provided for deeply understanding the proposed metric. By leveraging the NSS, we also proposed a skewness based dataset robustness metric for evaluating a DNN's adversarial performance on a given dataset. Extensive experiments using widely used state of the art architectures along with popular classification datasets, such as MNIST, CIFAR-10, CIFAR-100, and ImageNet, are used to validate the effectiveness and generalization of our proposed metrics. Instead of simply measuring a DNN's adversarial robustness in the input domain, as previous works, the proposed NSS is built on top of insightful mathematical understanding of the adversarial attack and gives a more explicit explanation of the robustness.\footnote{\textbf{Codes for the proposed metric and experiments, for further research and development, will be available upon acceptance.}}

\end{abstract}

\begin{figure*}[ht]
        \begin{subfigure}[b]{0.18\textwidth}
                \includegraphics[width=\linewidth, height=30mm]{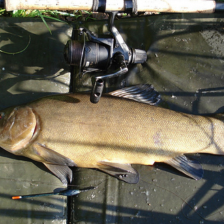}
                \caption{GT: Tench}
                \label{fig:1_org}
        \end{subfigure}%
        \hspace{0.5em}
        \begin{subfigure}[b]{0.18\textwidth}
                \includegraphics[width=\linewidth, height=30mm]{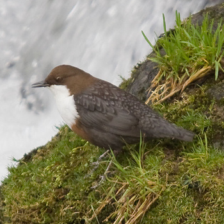}
                \caption{GT: Water ouzel}
                \label{fig:2_org}
        \end{subfigure}%
        \hspace{0.5em}
        \begin{subfigure}[b]{0.18\textwidth}
                \includegraphics[width=\linewidth, height=30mm]{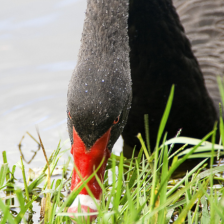}
                \caption{GT:Black swan}
                \label{fig:3_org}
        \end{subfigure}%
        \hspace{0.5em}        
        \begin{subfigure}[b]{0.18\textwidth}
                \includegraphics[width=\linewidth, height=30mm]{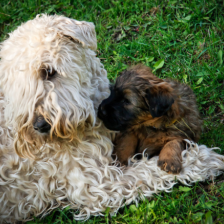}
                \caption{\makecell{GT: Wheaten terrier}}
                \label{fig:4_org}
        \end{subfigure}%
        \hspace{0.5em}        
        \begin{subfigure}[b]{0.18\textwidth}
                \includegraphics[width=\linewidth, height=30mm]{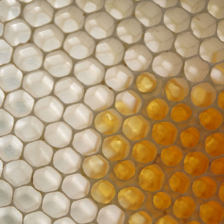}
                \caption{GT: Honeycomb}
                \label{fig:5_org}
        \end{subfigure}
        \vspace{1em}\\        
        \begin{subfigure}[b]{0.18\textwidth}
                \includegraphics[width=\linewidth, height=30mm]{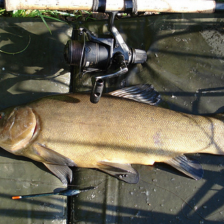}
                \caption{\makecell{MC: Coho\\ $\mu=0.00117, \epsilon=0.0007$}}
                \label{fig:1_ad}
        \end{subfigure}%
        \hspace{0.5em}        
        \begin{subfigure}[b]{0.18\textwidth}
                \includegraphics[width=\linewidth, height=30mm]{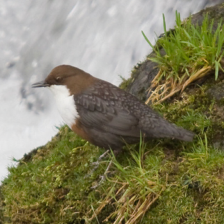}
                \caption{\makecell{MC: Partridge\\ $\mu=0.00156, \epsilon=0.0015$}}
                \label{fig:2_ad}
        \end{subfigure}%
        \hspace{0.5em}        
        \begin{subfigure}[b]{0.18\textwidth}
                \includegraphics[width=\linewidth, height=30mm]{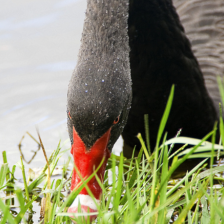}
                \caption{\makecell{MC: European gallinule\\ $\mu=0.00195, \epsilon=0.0017$}}
                \label{fig:3_ad}
        \end{subfigure}%
        \hspace{0.5em}        
        \begin{subfigure}[b]{0.18\textwidth}
                \includegraphics[width=\linewidth, height=30mm]{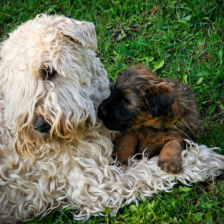}
                \caption{\makecell{MC: Irish terrier\\ $\mu=0.00293, \epsilon=0.0022$}}
                \label{fig:4_ad}
        \end{subfigure}%
        \hspace{0.5em}        
        \begin{subfigure}[b]{0.18\textwidth}
                \includegraphics[width=\linewidth, height=30mm]{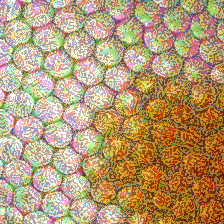}
                \caption{\makecell{MC:chainlink fence\\ $\mu=0.39063, \epsilon=0.1482$}}
                \label{fig:5_ad}
        \end{subfigure}        
        \centering
        
\caption{An illustration of the effect of FGSM attack on a pre-trained DenseNet-161 model. The images from (a)-(e) are the original images with their respective Ground Truth (GT) labels, and (f)-(j) are the crafted adversarial images along with their respective Mis-Classified (MC) labels. The corresponding maximum allowable noise level ($\mu$) and the used noise level ($\epsilon$) used by the FGSM are also indicated.}
		\label{fig:adversarial}
\end{figure*}
\IEEEpeerreviewmaketitle
\section{Introduction}
In recent years, machine learning and in particular deep learning has impacted various different fields, such as computer vision, natural language processing, and sentiment analysis. We have witnessed algorithms achieving human-like performance on a number of highly competitive datasets. The advent of deep learning has revolutionized the field of computer vision. The one interesting observation through all these results is the difference between machine and human vision. This difference is becoming more prominent after robustness flaws were found in neural network architectures \cite{szegedy2013intriguing}. Despite their outstanding performances, these algorithms tend to blow up when tested with adversarial inputs: perturbed inputs to force an algorithm to provide adversary-selected outputs. Adversarial machine learning is a branch of research that lies between machine learning and computer security. The vulnerability of deep networks to adversarial inputs is getting significant attention as they are used in various security and human safety applications. In general, an adversarial sample is generated by minimal perturbations of input data that are imperceptible to a human observer. More importantly, these adversaries are shown to be transferable between different architectures \cite{NIPS2014_5423}. Exploring this transferability property, it was observed that adversaries can be created using black-box attacks by training an unknown architecture on similar data \cite{papernot2017practical}. Besides black-box attacks, there are many other adversarial attack schemes. These schemes can be divided into two categories, namely fix-directional and unfix-directional attack. The direction in the context of adversarial attack means the direction of the generated adversarial vector used for perturbing a specific input, which means the direction is input-wise. During an adversarial attack, if the direction of the generated vector keeps unchanged, it falls into the fix-directional attack, and vice-versa. In this paper, we mainly focus on fix-directional attacks.

One would expect state-of-the-art architectures to be robust in these adversarial environments, but our experimental results do not confirm that. For example, the pre-trained DenseNet-161 from PyTorch~\cite{paszke2017automatic} gives fairly high top-1 validation accuracy $77.14$\% on ImageNet Large Scale Visual Recognition Challenge (ILSVR) 2012 dataset. Surprisingly, it falls to $35.2$\% under a Fast Gradient Sign Method attack with an allowable noise level up to $\mu=0.00117 (0.3/256)$. With an increase in noise level, the validation accuracy decreases further. When the allowable noise level increases up to $\mu=0.00293 (0.75/256)$, the validation accuracy of DenseNet-161 is only $15.19$\%. In Fig.~\ref{fig:adversarial}, five original and misclassification adversarial images are illustrated. The adversarial images are obtained from a FGSM attack to the DenseNet-161. As shown in Fig.~\ref{fig:adversarial}, there is no noticeable perceptual difference among the first four image pairs (e.g., Fig.~\ref{fig:1_org} vs Fig.~\ref{fig:1_ad}) which makes the adversarial attacks significantly dangerous. This is mainly because the allowable noise level is controlled in a low value range (i.e., up to $\mu=0.00293$). We can also give a very high allowable noise level, say $\mu=0.39063 (100/256)$, as shown in Fig.~\ref{fig:5_org} and Fig.~\ref{fig:5_ad} in order to observe the noticeable noise in the adversarial image. Thus, an ideal adversarial attack needs to control the allowable noise level in order not to bring any noticeable perceptual difference. 

Therefore, an extensive adversarial performance evaluation is needed for all DNNs that are to be used for real-world applications. Relying solely on classification accuracies is not sufficient to define the robustness of an architecture. Even though many adversarial robustness metrics are proposed, most of them are only validated on either very limited small datasets (e.g., MNIST and CIFAR-10) or with very limited DNN models, which makes the proposed metrics unconvincing, especially in terms of the metrics' scalability \cite{papernot2016limitations,tabacof2016exploring,bastani2016measuring,carlini2017towards}. A scalable and more convincing robustness metric is still missing from the literature and we aim to bridge this gap using our proposed work. In this paper, a novel Noise Sensitivity Score (NSS) metric is proposed, which is able to quantize the adversarial robustness of a DNN's input under a fix-directional adversarial attack. We model fix-directional adversarial attacks mathematically and provide an intuitive explanation about why and how does the proposed NSS work, which is also missed by most of the previous works. Based on the proposed NSS, a skewness based dataset robustness metric is also proposed to evaluate a DNN's adversarial performance on a specific dataset. More importantly, the proposed NSS and skewness based dataset robustness metric are extensively tested with different state-of-the-art DNN models (e.g., DenseNet-161, ResNet-152) across different widely used classification datasets (i.e., MNIST, CIFAR-10, CIFAR-100, ImageNet), which gives a potential user more confidence to trust the proposed metrics. This paper makes the following contributions:
\begin{itemize}
\item We intuitively and mathematically explain a DNN's fix-directional adversarial attack and propose NSS and skewness based dataset robustness metric on top of it. The expandability of the proposed metrics are vital in the DNN society.
\vspace{0.25em}
\item We present a diverse set of experimental results, and show that a) our proposed robustness metrics are effective and generalizable on different DNNs, datasets, and attack types. b) the skewness based dataset robustness metric provides a more intuitive understanding than adversarial accuracy. 
\end{itemize}


This paper is organized as follows. We summarize the background to our paper in Section \ref{sec:background}. In Section \ref{sec:method}, we elaborate on the proposed robustness metric score by providing an extensive explanation to understand it's role in adversarial performance. Results from comparative experiments for different architectures and datasets are given in Section \ref{sec:evaluation}. We then discuss various insights that we gained from our current work in section \ref{sec:discussion}. We provide a brief summary of the related works in section \ref{sec:relwork}. Conclusions are finally drawn in Section \ref{sec:conclusion}.

\vspace{-2mm}
\section{Background}
\label{sec:background}
Adversaries in neural networks have become an area of active research after experiments showed that neural networks are prone to attacks from adversaries \cite{szegedy2013intriguing}. Additionally, the use of neural networks in critical applications, such as security, surveillance systems, and medical imaging, requires a high degree of robustness from these models. With regards to adversaries in deep learning, research is simultaneously done in three directions. First, given an input distribution, crafting adversarial samples using novel adversarial algorithms. Second, proposing methods to develop defenses against these adversaries, and last, introducing metrics for measuring the robustness of a DNN.
\subsection{Adversarial Attacks}
\label{related:1}
Over the past few years, extensive work has been done in devising new methods to generate adversarial samples. 
Broadly, adversarial attacks can be divided into gradient based attacks, non-gradient based attacks and black-box attacks. Gradient attacks are further sub-divided into fix-directional and unfix-directional attacks. Fix-directional gradient attacks have been extensively used in the literature \cite{xu2017feature,papernot2017practical}. Szegedy et al. \cite{szegedy2013intriguing} proposed L-BFGS algorithm to craft adversarial samples and showed that these samples possess transferability property. 
Goodfellow et al. proposed Fast Gradient Sign Method (FGSM) - a fast approach  for generating adversarial samples by adding perturbation proportional to the sign of the cost functions gradient \cite{goodfellow2014explaining}. FGSM as defined, moves one in the direction (fixed) of the border between the true class and some other class. Rather than adding perturbation throughout the image, Pappernot el al. \cite{papernot2016limitations} proposed  Jacobian Saliency Map Approach (JSMA) - the notion of adversarial saliency maps in which only the most sensitive input components were only perturbed.

Kurakin et al. proposed Iterative Gradient Sign Method (IGSM) as a simple extension to FGSM \cite{kurakin2016adversarial}. Instead of applying the adversarial noise once, the method applies it iteratively with smaller noises. Additionally, in each iterative step the direction of the attack could change leading to a unfix-directional attack. Evidently, IGSM is cost ineffective as compared to FGSM attack \cite{xu2017feature}. DeepFool computed and applied the minimal perturbation necessary for misclassification under the L2-norm \cite{moosavi2016deepfool}. In this approach, the algorithm performs iterative steps in the adversarial direction of the gradient provided by a local linear approximation of the classifier. As a result, the approximation is more accurate than FGSM and faster than JSMA, as all the pixels are simultaneously modified at each step of the method. As an extension of this work, Moosavi-Dezfooli et al. crafted an universal perturbation  indifferently applicable to any instance \cite{moosavi2017analysis}. The perturbation is computed in a greedy fashion and needs multiple iterations for convergence. Carlini et al. proposed an efficient method to compute good approximations while keeping the computational cost of perturbing examples low \cite{carlini2017towards}. It defined three similar targeted attacks, based on different distortion measures: $L_{0}$, $L_{2}$, and $L_{\infty}$ respectively.

It is to be noted that all the above mentioned attacks falls under the general category of gradient attacks. 
In the classical black-box attack, the adversarial algorithm has no knowledge of the architectural choices made to design the original architecture, which includes the number, type, and size of layers, nor of the training data used to learn the architectural hyper-parameters. The algorithm's only capability is to observe the labels assigned by the network for chosen inputs \cite{papernot2017practical}. There are different ways to generate adversarial samples using black-box schemes. Detailing all of them is outside the scope of this paper.

\subsection{Adversarial Defenses}\label{related:2}

Notably, it is difficult to find new methods that are effective in threatening a model, and even tougher to defend an algorithm from it. There have been some suggestions in the literature to design defensive mechanisms for deep learning architectures. Goodfellow et al. proposed adversarial training in which an algorithm is retrained with adversarial images alongside the legitimate dataset \cite{goodfellow2014explaining} . Pappernot et al. extended defensive distillation---which is one of the mechanisms proposed to mitigate adversarial examples---to address its limitations \cite{papernot2017extending}. They revisited the defensive distillation approach and used soft labels to train the distilled model. The resultant model was robust to attacks, such as FGSM and JSMA. 
Liang et al. proposed a method where the perturbation to the input images are regarded as a kind of noise and the noise reduction techniques are used to reduce the adversarial effect \cite{liang2017detecting}. Classical image processing operations, such as scalar quantization and smoothing spatial filters were used to reduce the effect of perturbations. Bhagoji et al. proposed dimensionality reduction as a defense against evasion attacks on different machine learning classifiers \cite{bhagoji2017dimensionality}. All the above mentioned works propose defensive mechanisms for a particular attack type. Clearly, scheming robust models against adversarial samples remains an open problem.

\subsection{Robustness Metrics}\label{related:3}
With the development of several adversarial attacks and defenses, there is a need for a robustness metric which quantifies the performance of a DNN against adversarial samples. Analogously, little work has been done in devising these metrics. Most works use adversarial accuracy, performance of a model in classifying generated adversarial samples, as the metric for determining the robustness of a model. Arguably, adversarial accuracies provides very little information concerning the amount of perturbation needed for attacking the network. Pedro et al. used the fraction of images (in \%) that keep or switch labels when noise is added to a departing image \cite{tabacof2016exploring}. This was used as a stability measure for the departing image in the pixel space of a classifier. This fraction was calculated over a sample of 100 probes, where each probe was a repetition of the experiment with all the factors held fixed but the sampling of the random noise. It is not efficient to use adversarial accuracy as a measure of robustness as they do not shed light on the general adversarial behavior of an algorithm at higher noises. 
Goodfellow et al. proposed adversarial training to improve the robustness of a neural network ($f$) by augmenting the training examples with the generated adversarial examples, and then train a new neural network ($f'$) \cite{goodfellow2014explaining}. Robustness is then evaluated by using the same adversarial algorithm to generate adversarial samples for $f'$. If the algorithm detects fewer adversarial samples, then $f'$ is concluded to be more robust than $f$. Adversarial training, in general, increases the adversarial performance of an algorithm. Measuring robustness in this way is  unproductive, as the overall robustness is directly proportional to the adversarial images used for retraining the network. We expect a higher robustness if we generate adversarial images with varying noises. Moreover, this metric is more data driven as the network may overfit to the adversarial samples. Various distance measures have been proposed to quantify the amount of perturbation added in generating adversarial samples. For instance, $L_{0}$ distance measured the number of pixels that have been altered in an image. This is easily arguable as the distance only counts the number of altered pixels and does not provide any knowledge with regards to the measure of perturbation. $L_{2}$ distance measures the standard root-mean-square distance between an image and its adversarial counterpart. It has been found that very small amount of perturbation is sufficient to generate adversarial samples. Evidently, the $L_{2}$ distance can remain very small even when many small changes are made throughout the image. Recently, $L_{\infty}$ has been argued as the optimal robustness metric \cite{warde201611}, measuring the maximum change to any of the pixels. These metrics are not validated for different attacks and architectures. Using point-wise robustness, Bastani et al. proposed two statistics, namely adversarial frequency and adversarial severity, for measuring robustness \cite{bastani2016measuring}. For efficiency purposes, the authors fixed the adversarial label to be the second most probable label for all the adversarial samples. One would expect that the adversarial sample for an image $X$ can be easily created using it's second most probable class. This assumption is more intuitional than practical, as many times the most efficient gradient direction is not in the direction of the label with second highest score. Also, the proposed adversarial severity metric does not work with signed gradient algorithm since all the adversarial examples it finds using $L_{\infty}$ had norm equal to the introduced noise level, $\epsilon$. Pappernot et al. defined the robustness of a DNN as its capability to resist perturbations \cite{papernot2016distillation}.  The intuition behind this metric is that a robust algorithm will ensure that its classification output will remain constant in a closed neighborhood for any sample taken from its input distribution. The larger this neighborhood is for all input samples, the more robust is the DNN. This approach only tests individual data points from an infinite input space and certainly does not carry forward to all other points. Also, the neighborhood value is different for each input points. One may always observe some input points closer to the classification boundary having a very small neighborhood as compared to the one further from it. To sum up, we believe a good robustness metric is vital in understanding the adversarial performance of a DNN.

\section{Approach}
\label{sec:method}
\subsection{Background}
\textbf{Deep Neural Network Adversarial Attack} A Deep Neural Network (DNN) $D: X \rightarrow Y$ can be regarded as an approximation of a differentiable non-linear function $F: X \rightarrow Y'$ which takes a vector $X \in R^N$ as input and generates an output vector $Y \in R^K$, where $Y$ is expected to be a close approximation of $Y'$. For each element of the output vector $y_i \in Y$, $D(X)$ defines an unique output surface $m_i$ for the input domain $X$. When $D(X)$ approximates a loss function $F(X)$ of a classifier, a single point $y_{i}$ on a output surface $m_i$ represents an error value. Specifically, for an input $X$, we have:
\begin{align}
\label{eq:cl}
D(X) &= t \nonumber \\
min(y_i) &= y_t \quad i \in 1\ldots K,
\end{align}
where $t$ is the predicted class label. A DNN misclassification adversarial attack with an input $X$ can be defined as:
\begin{equation}
\label{eq:ad}
\underset{\delta{x}}{\mathrm{argmin}}\|\delta{x}\| \quad s.t. \quad D(X + \delta{x}) \neq D(X),
\end{equation}
where $\|\cdot\|$ is a appropriate norm metric in the input domain ($L_2$ norm is used in our experiments), and $\delta{x}$ is the smallest perturbations of $X$ used for satisfying the constraint of Eq.~\eqref{eq:ad}. Intuitively, Eq.~\eqref{eq:ad} means that the smallest perturbation $\delta{x}$ moves the original point $y_{i}$ to a new position $y'_{i}$ on the output surface $m_i$ which disqualifies Eq.~\eqref{eq:cl}. Many attack approaches~\cite{goodfellow2014explaining, kurakin2016adversarial, moosavi2016deepfool} have been proposed to find the most effective $\delta{x}$ efficiently. In this paper, we focus on fix-directional adversarial attack methods in which the $\delta{x}$ is defined as: 
\begin{equation}
\label{eq:ga}
	\delta{x} = \epsilon \vec{V}, 
\end{equation}
where $\vec{V}$ is a generated adversarial vector. FGSM is one of the fix-directional attacks, where $\delta{x}$ is defined as:
\begin{equation}
\label{eq:gas}
	\delta{x} = \epsilon \vec{V}_{\text{FGSM}} =\epsilon\text{sign}\frac{\partial{y_t}}{\partial{X}}. 
\end{equation}
Specifically, as shown in Eq.~\eqref{eq:gas}, $\vec{V}_{\text{FGSM}}$ is the gradient vector generated by backpropagating the error $y_t$ with respect to the input $X$. In order to obtain the most effective $\delta{x}$, a fix-direction attack only changes the value of $\epsilon$ (i.e., noise level) while keeping the $\vec{V}$ fixed. Contrastively, Iterative Gradient Sign Method (IGSM)\cite{kurakin2016adversarial} is not a fix-directional attack. It is because, in each iterative step, the direction of the gradient vector could be changed.\vspace{0.5em}

\textbf{Data Jacobian Matrix} We denote the Data Jacobian Matrix (DJM) of a DNN for an input $X$ as:
\begin{equation}
\label{eq:DJM}
	DJM(X) = \frac{\partial{F(X)}}{\partial{X}} =
	\begin{bmatrix}
    \frac{\partial{F_i(X)}}{\partial{x_j}}  
\end{bmatrix}_{i\in 1 \ldots K, j\in 1 \ldots N}
\end{equation}
Essentially, $DJM(X)$ is very similar to the gradient backpropagated through a DNN during a training process. The only difference is $DJM(X)$ differentiates with respect to the input features rather than network parameters. Since $F(X)$ is differentiable at a point $y_i$ on the surface $m_i$, the $i$th row of the $DJM(X)$, denoted as $DJM_i(X)$, defines a linear transform $R^N \rightarrow R^1$, which is the best pointwise linear approximation of the function $F(X)$ near the point $y_i$ on the surface $m_i$~\cite{wikiJM}, which can be mathematically represented as:
\begin{equation}
\label{eq:DJM_LT}
	F(X+\delta{x}) = F(X) + DJM(X)\times\delta{x} + \delta{e},
\end{equation}
where $\delta{e} \in R^K$ is the difference between original surfaces and approximated surfaces. 

\subsection{Sensitivity Skewness Robustness Metric}
\label{sec:SRM}
Local neighborhoods on a output surface around different surface points have different shapes, which results in different inputs have different sensitivity to the same noise level $\epsilon$. In order to measure the sensitivity, we propose a novel Noise Sensitivity Score (NSS) as an input-wise adversarial robustness metric. The core of the proposed NSS is a novel $DJM$ based rate of change estimation. They are explained from an insightful mathematical point of view in section~\ref{sssec:NSM} and ~\ref{sssec:cf} respectively. For a certain dataset, we are able to measure the robustness of a DNN to the dataset under a fix-directional attack based on the skewness of the obtained NSSs distribution, which is illustrated in section~\ref{sssec:DSM}.\vspace{0.25em} 

\subsubsection{Noise Sensitivity Score (NSS)}
\label{sssec:NSM}
Without loss of generality, let $D(X)$ approximate a loss function of a classifier as before. With an allowable maximum noise level $\mu$, a successful misclassification adversarial attack in this context is the added noise $\delta{x} \le \mu$ disqualifies the Eq.~\eqref{eq:cl}, which gives:
\begin{equation}
\label{eq:mca}
	min(y_i) \neq y_t \quad i \in 1\ldots K.
\end{equation}
For an unsuccessful attack, the Eq.~\eqref{eq:cl} still holds. 
Since the current added noise $\delta{x}$ is not able to deliver a successful misclassification attack, a fix-directional attack increases the noise level $\epsilon$ up to the allowable maximum noise level $\mu$ in order to deliver a successful misclassification attack. The proposed NSS is able to reflect the noise sensitivity of a specific input within a controlled enlarged noise level, which is defined as:
\begin{align}
\label{eq:s_score}
	NSS& :=  \frac{y_j - y_t}{s_t - s_j} \\
    \quad &\textrm{s.t.} \quad\underset{j}{\mathrm{argmin}}(f(\frac{y_j - y_t}{s_t - s_j}))\quad j \in 1\ldots K, j\neq t, \nonumber \\
    &\textrm{where, } f(x) = \begin{cases} C, & x < 0 \\ 
    x, & x  \geqslant 0  \end{cases} \nonumber 
\end{align}
where $y_j - y_t$ is the gap needed to be covered in order to misclassify the input to the $j$th class. Since we only consider the correctly classified inputs, the value of $y_j - y_t$ is always positive. The $s_i$ is the rate of change on $y_i$ dimension which is driven by the added noise $\delta{x}$. The minus sign in the denominator is a bit tricky, which is because an ideal misclassification attack should increase the target error value $s_t$ and decrease the error value of the potential misclassfied class $s_j$, which means the value of $s_t - s_j$ should be positive. However, the $s_t - s_j$ could be a negative value in the following situations: (a) $s_t$ decreases and $s_j$ increases; (b) both of $s_t$ and $s_j$ decrease, but $s_t$ decreases faster; (c) both of $s_t$ and $s_j$ increase, but $s_j$ increases faster. If any of the above cases happen to an input, it means it is impossible for a fix-directional attack to deliver an effective attack to the input. Then, the specific input deserve a maximum score. This task is achieved by the $f(x)$ is used in the Eq.~\eqref{eq:s_score}. The value of $C$ defined in $f(x)$ is a constant positive value the so-called maximum score. In section~\ref{sssec:DSM}, we show that $C$ could be any value larger than the predefined skewness threshold. Intuitively, Eq.~\eqref{eq:s_score} can be regarded as finding the minimum time to cover the distance $y_j - y_t$ with a speed $s_t - s_j$. Therefore, the input with a smaller NSS is more vulnerable to the enlarged noise, and vice versa.


A local neighborhood on an output surface $m_i$ is rarely flat, especially for an enlarged local area resulted from an increased noise level $\epsilon$. This raises the difficulties to estimate the rate of change in the level of desired accuracy. Therefore, we propose to estimate the rate of change based on a novel $DJM$ based scheme, which takes the nonlinearity nature of an output surface into account. \vspace{0.25em}

\subsubsection{Estimating Rate of Change}
\label{sssec:cf}
The essential idea of the proposed $DJM$ based scheme is leveraging the aforementioned feature of $DJM$ which states that $DJM$ gives the best pointwise linear approximation of a function in a small local neighborhood around a point in a embedded space. Specifically, if an added noise $\delta{x}$ does not deliver a successful attack, based on the Eq.~\eqref{eq:DJM_LT}, with a small extra shift $\delta{x'}$ in the input domain, the approximated shifted position of a surface point $y_i^n$ can be estimated as:
\begin{equation}
\label{eq:DJM_tan}
	y_i' \approx F(X + \delta{x}) + DJM_i(X + \delta{x})\times\delta{x'}. 
\end{equation}
Therefore, the rate of change caused by the small shift $\delta{x'}$ can be estimated as:
\begin{equation}
\label{eq:DJM_speed}
	s_i \approx \frac{DJM_i(X + \delta{x})\times\delta{x'}}{\|\delta{x'}\|}\quad i \in 1\ldots K.
\end{equation}
It is worth to mention that, the $\delta{x}$ in the above equation can be of $0$, which means an attack starts from the original input $X$ instead of $X + \delta{x}$. This feature gives the proposed NSS more flexibility to be used. The validation results in section~\ref{sssec: prop} illustrate the effectiveness of the proposed NSS and the $DJM$ based rate-of-change estimation. However, since the $DJM$ is only an accurate linear approximation for a small $\delta{x'}$. The approximate accuracy decreases while $\delta{x'}$ increases. This can also be observed in our validation results. More details related to the validation are provided in section~\ref{sssec: prop}.\vspace{0.25em}

\subsubsection{Skewness Based Dataset Robustness Metric}
\label{sssec:DSM}
The proposed NSS is an input-wise adversarial robustness estimator. In order to understand the adversarial performance of a DNN on a dataset, we need to obtain the NSSs for all correctly classified inputs. Intuitively, if more inputs give smaller NSSs, it means the DNN is less robust to the attack on the dataset, and vice versa. This phenomenon can be quantitatively described by the skewness of the obtained NSSs' distribution. Due to the perceptual constraint, with a controlled noise level, a fix-directional attack is unable to successfully attack all the inputs of a dataset. It means only the distribution of smaller NSSs matters to the dataset robustness score. Thus, a NSS threshold is introduced and the skewness is estimated only based on the NSSs that are smaller than the defined threshold. Empirically, the threshold is set to 5 for all our tested datasets. That also explains the reason that the value of $C$ defined in Eq.~\eqref{eq:s_score} can be of any value larger than the NSS threshold. In other words, the proposed robustness score is not sensitive to the value $C$. We set $C$ to 100 in our experiments in order to view the full picture of the obtained NSSs' distribution. To calculate the skewness, a commonly used sample skewness~\cite{kokoska2000crc} is adopted, which is defined as:
\begin{equation}
\label{eq:skewness}
	\gamma=\frac{\sqrt{n(n-1)}}{n-2}\frac{\frac{1}{n}\sum\limits_{i=1}^n(x_i - \bar{x})^3}{(\frac{1}{n}\sum\limits_{i=1}^n(x_i - \bar{x})^2)^\frac{3}{2}},
\end{equation}
where $x_i$ is the sample with NSS larger than the predefined threshold; $\bar{x}$ is the mean of those samples; $n$ is the number of those samples.

\section{Evaluation}
\label{sec:evaluation}

Our evaluations are designed to demonstrate the effectiveness and generalization of the proposed NSS and the skewness based dataset adversarial robustness metric. To do that, we apply the proposed metrics to different widely used state of the art DNN models on four different well known classification datasets, namely MNIST \cite{lecun1998mnist}, CIFAR-10 \cite{krizhevsky2009learning}, CIFAR-100 \cite{krizhevsky2009learning}, and ImageNet \cite{imagenet_cvpr09}. Without loss of generality, two widely known fix-directional attacks are used, which are Fast Gradient Sign Method (FGSM) and Fast Gradient Method (FGM). In section~\ref{sec:ad_crafting}, we first describe FGSM and FGM in more details. Then, all the chosen datasets and corresponding DNN models are described in section~\ref{sec:dataArch}. In section~\ref{sec:vali_results}, the validation results are illustrated in terms of NSS and skewness based dataset adversarial robustness.

\subsection{Adversarial Crafting}
\label{sec:ad_crafting}
The natural goal of an adversary is to craft adversarial samples. As discussed in Section \ref{related:1}, there are different ways of formulating such adversarial samples. For validation purpose, we mainly focus on fix-directional gradient attacks where we adopt FGSM and FGM specifically. Eq.~\eqref{eq:gas} gives the mathematical definition of FGSM. Similarly, FGM is mathematically defined as:
\begin{equation}
\label{eq:FGM}
	\delta{x} = \epsilon \vec{V}_{\text{FGM}} =\epsilon\frac{\partial{y_t}}{\partial{X}}.
\end{equation}
As we can see, the only difference between Eq.~\eqref{eq:gas} (i.e., FGSM) and Eq.~\eqref{eq:FGM} (i.e., FGM) is that FGSM applies the sign operator to the calculated gradient. The sign operator can be regarded as a normalization process, which results in a more perceptually plausible adversarial image. Theoretically, $\vec{V}_{\text{FGM}}$ gives the most effective direction to perform an adversarial attack. However, the sign operator used in FGSM has changes the most effective direction, which makes the FGSM is less effective than FGM. Our extensive experiment results also illustrate that FGM is a stronger attack as compared to FGSM. 

\subsection{Datasets and DNN Architectures}
\label{sec:dataArch}
\noindent\textbf{MNIST}.
Our initial experiments were performed on standard MNIST dataset of handwritten digits. Each sample is a $28\times28$ binary pixel image. The complete dataset consists of 70k images, divided into 60k training and 10k testing images. For preprocessing, we normalize the data using mean and standard deviation. Three different architectures are used on this dataset, namely, \textbf{LeNet} \cite{lecun1998gradient} and two three-layer Multiple Layer Perceptron(MLP). The only difference between the two MLPs is the number of neurons in each layer is set differently. One of them has 100 neurons in each layer, denoted by \textbf{MLP1}, while the other one has 200 neurons in each layer, which is denoted by \textbf{MLP2}.  All three networks are trained with a learning rate of $\eta = 0.01$ for 20 epochs by using Stochastic Gradient Descent (SGD).\vspace{0.25em}

\noindent\textbf{CIFAR}.
CIFAR dataset comprises of two sub-datasets, namely, CIFAR-10 and CIFAR-100. Both the CIFAR datasets consist of $32\times32$ RGB images. CIFAR-10 consists of 10 classes, and CIFAR-100 contains images drawn from 100 classes. Each of them contains 60k images, 50k training images, and 10k testing images. Preprocessing was done to normalize the data using channel means and standard deviation. Additionally, a widely used data augmentation scheme of mirroring and shifting are used for both CIFAR-10 and CIFAR-100. For validation, \textbf{VGG-13} \cite{simonyan2014very}, \textbf{ResNet-18} \cite{he2016deep} and \textbf{DenseNet-40-without-bottleneck} \cite{huang2017densely} models are used due to the  following reasons: a) All of them are widely used state of the art models, and b) They represent three different mainstream architectures, namely sequential, skip-connected and cross-connected. Therefore, the validations based on these three networks demonstrates the generality of the proposed metrics. CIFAR-10 and CIFAR-100 architectures are trained for 300 epochs respectively. A non-uniform learning rate scheme is used, where the initial learning rate is set to 0.1, and is divided by 10 at 50\% and 75\% of the total number of epochs.\vspace{0.25em}

\noindent\textbf{ImageNet}.
ImageNet Large Scale Visual Recognition Challenge (ILSVR) 2012 consisting of 1.2 million training images and 50,000 validation images is used for our validation. With the same motivation as CIFAR, four best performed pre-trained models provided by PyTorch \cite{paszke2017automatic} are chosen for our validation purpose, which are \textbf{VGG-19 with batch normalization}, \textbf{Inception-v3}, \textbf{ResNet-152}, and \textbf{DenseNet-161}. All of the mentioned networks expect the input images to be normalized in mini-batches of 3-channel RGB images of shape 

\begin{figure*}[ht]
        \begin{subfigure}[b]{0.33\textwidth}
                \includegraphics[width=\linewidth, height=30mm]{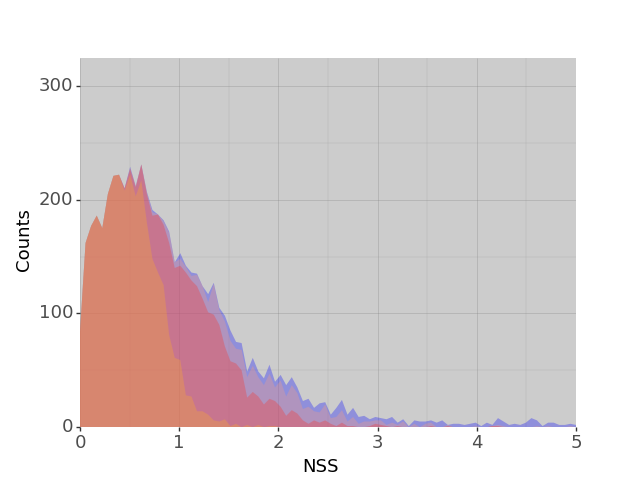}
                \caption{MNIST-MLP1}
                \label{fig:MNIST-MLP1-FGSM}
        \end{subfigure}%
        \begin{subfigure}[b]{0.33\textwidth}
                \includegraphics[width=\linewidth, height=30mm]{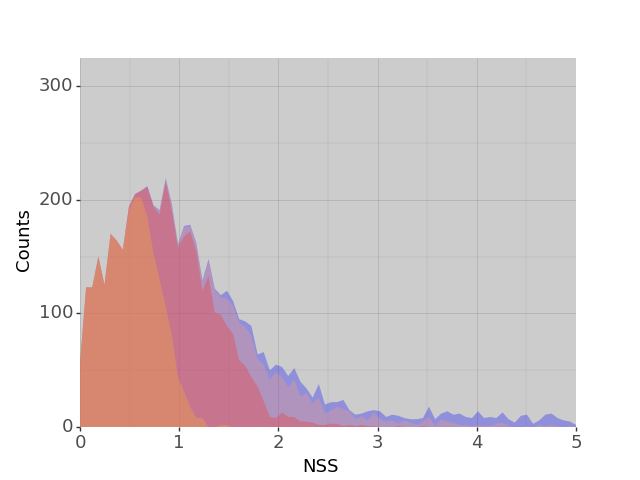}
                \caption{MNIST-MLP2}
                \label{fig:MNIST-MLP2-FGSM}
        \end{subfigure}%
        \begin{subfigure}[b]{0.33\textwidth}
                \includegraphics[width=\linewidth, height=30mm]{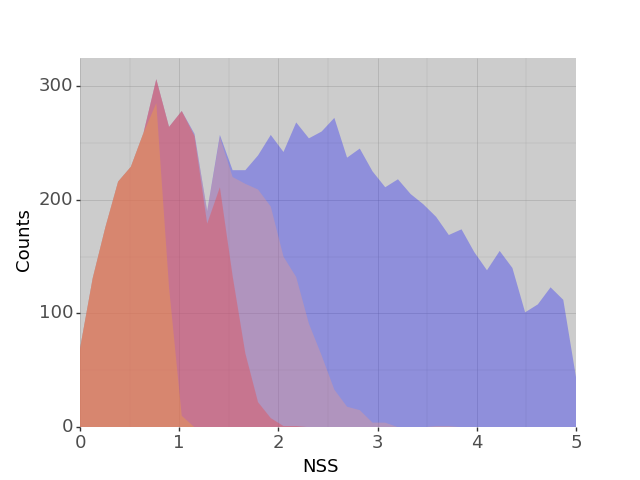}
                \caption{MNIST-LeNet}
                \label{fig:MNIST-LENET-FGSM}
        \end{subfigure}
        \begin{subfigure}[b]{0.33\textwidth}
                \includegraphics[width=\linewidth, height=30mm]{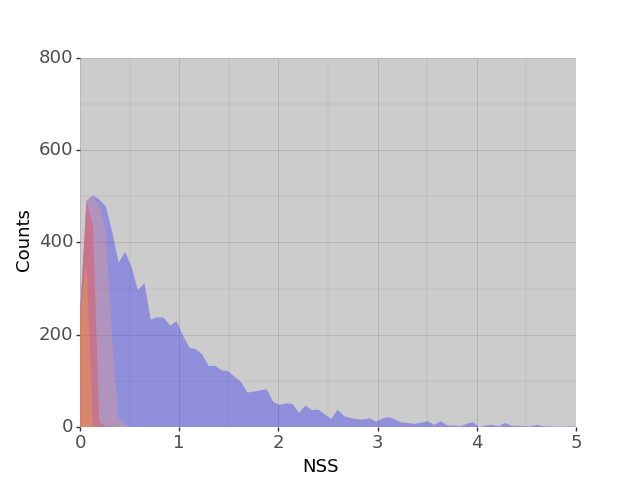}
                \caption{CIFAR10-DenseNet-40 w/o Bottleneck}
                \label{fig:CIFAR10-DenseNet-FGSM}
        \end{subfigure}%
        \begin{subfigure}[b]{0.33\textwidth}
                \includegraphics[width=\linewidth, height=30mm]{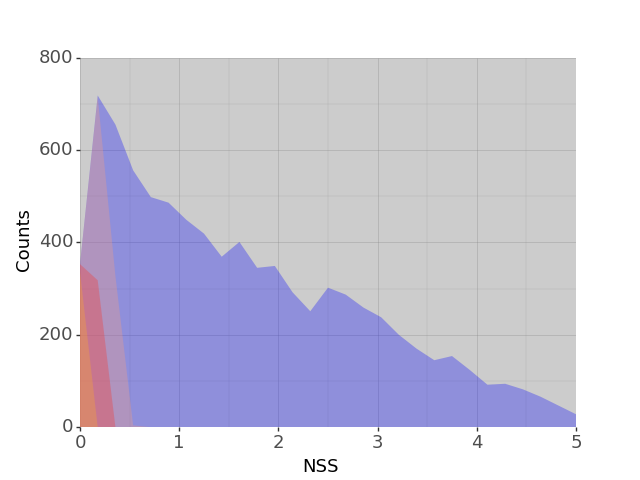}
                \caption{CIFAR10-ResNet18}
                \label{fig:CIFAR10-Res18-FGSM}
        \end{subfigure}%
        \begin{subfigure}[b]{0.33\textwidth}
                \includegraphics[width=\linewidth, height=30mm]{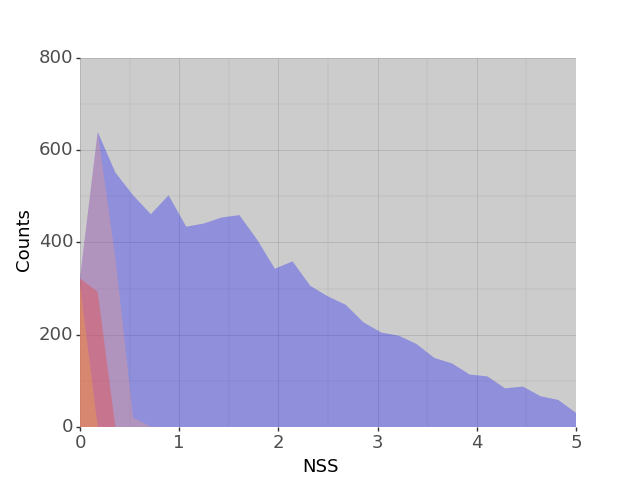}
                \caption{CIFAR10-VGG13}
                \label{fig:CIFAR10-VGG13-FGSM}
        \end{subfigure}
        \begin{subfigure}[b]{0.33\textwidth}
                \includegraphics[width=\linewidth, height=30mm]{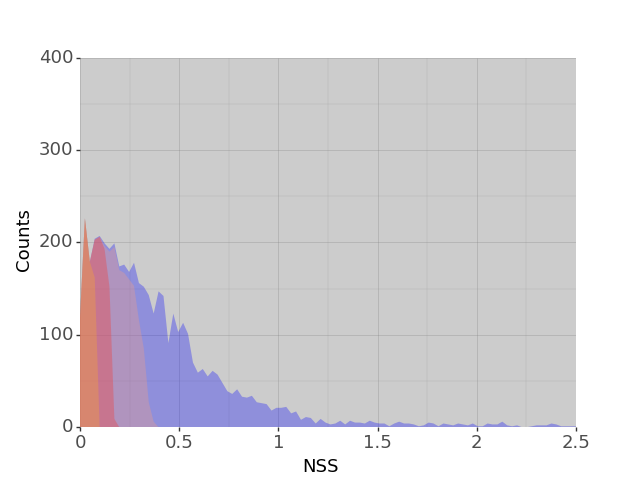}
                \caption{CIFAR100-DenseNet-40 w/o Bottleneck}
                \label{fig:CIFAR100-DenseNet-FGSM}
        \end{subfigure}%
        \begin{subfigure}[b]{0.33\textwidth}
                \includegraphics[width=\linewidth, height=30mm]{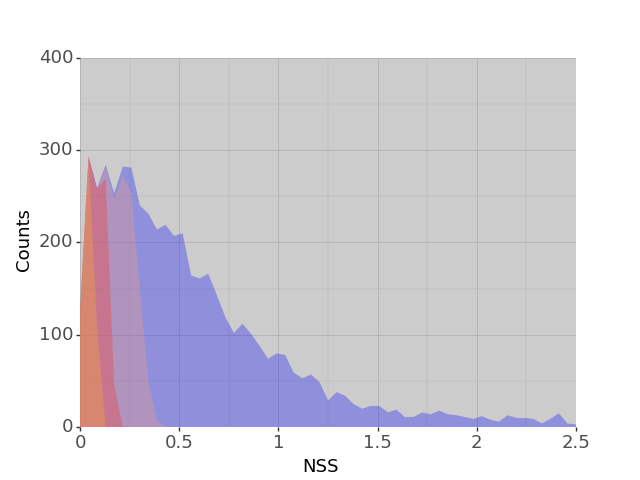}
                \caption{CIFAR100-ResNet-18}
                \label{fig:CIFAR100-Res18-FGSM}
        \end{subfigure}%
        \begin{subfigure}[b]{0.33\textwidth}
                \includegraphics[width=\linewidth, height=30mm]{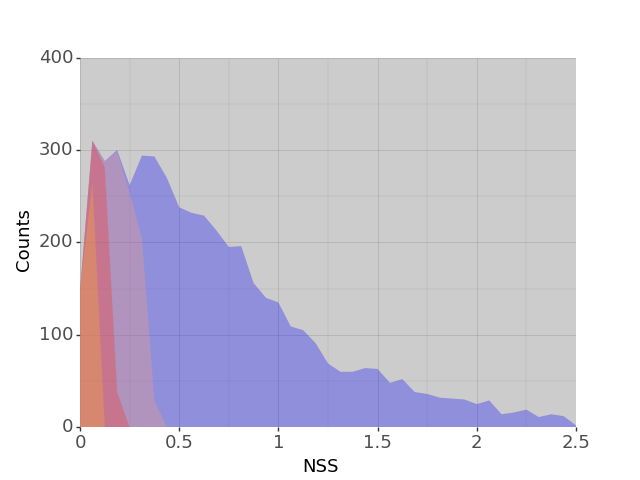}
                \caption{CIFAR100-VGG13}
                \label{fig:CIFAR100-VGG13-FGSM}
        \end{subfigure}
        \begin{subfigure}[b]{0.25\textwidth}
                \includegraphics[width=\linewidth, height=30mm]{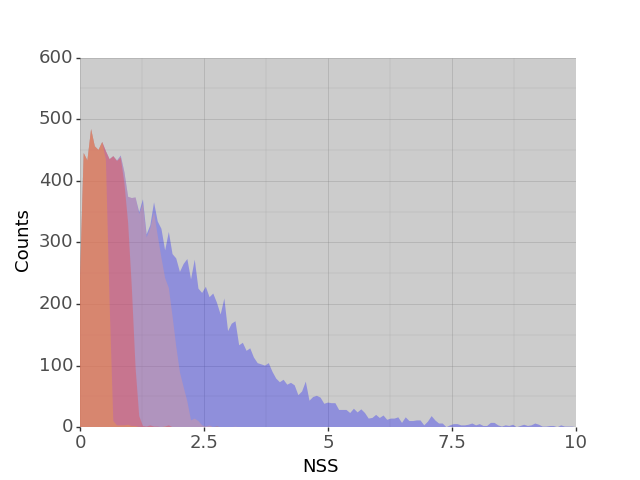}
                \caption{IMAGENET-DenseNet-161}
                \label{fig:IMAGENET-DENSENET161-FGSM}
        \end{subfigure}%
        \begin{subfigure}[b]{0.25\textwidth}
                \includegraphics[width=\linewidth, height=30mm]{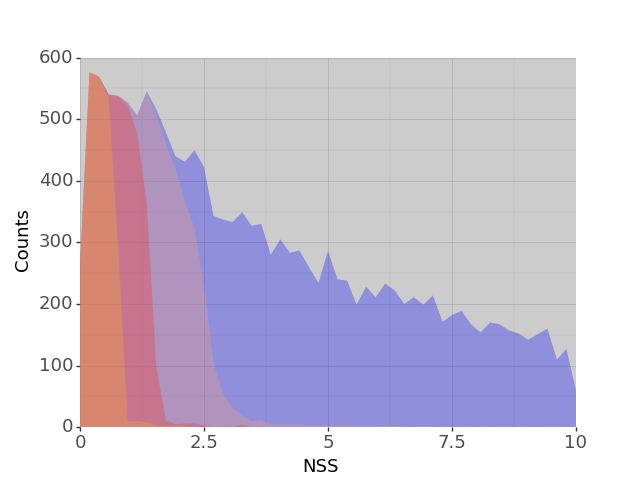}
                \caption{IMAGENET-Inception-V3}
                \label{fig:IMAGENET-INCEPTION-FGSM}
        \end{subfigure}%
        \begin{subfigure}[b]{0.25\textwidth}
                \includegraphics[width=\linewidth, height=30mm]{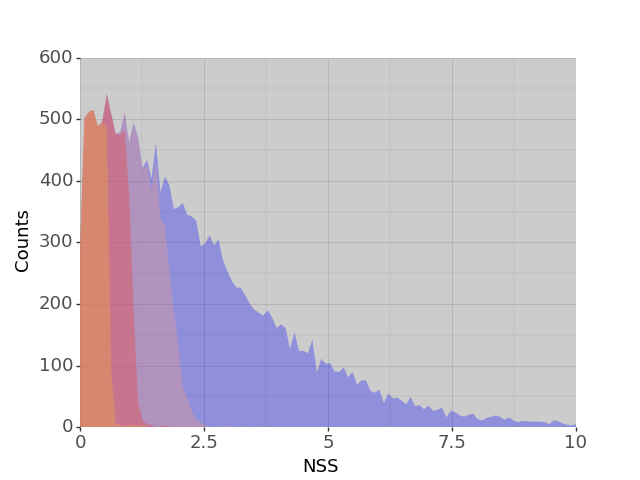}
                \caption{IMAGENET-ResNet-152}
                \label{fig:IMAGENET-RESNET-152-FGSM}
        \end{subfigure}%
        \begin{subfigure}[b]{0.25\textwidth}
        \centering
                \includegraphics[width=\linewidth, height=30mm]{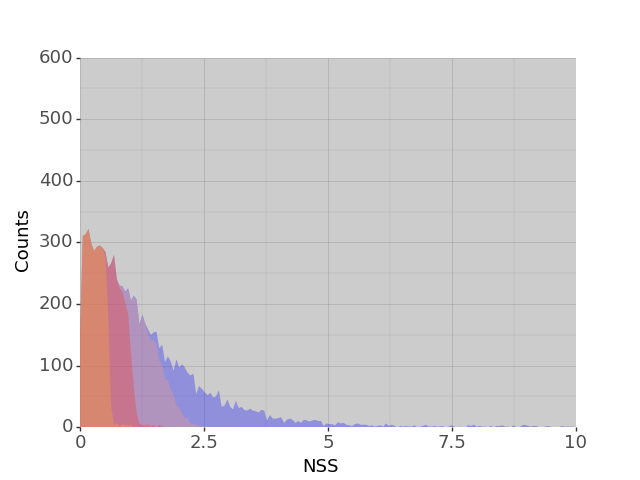}
                \caption{IMAGENET-VGG-19-BN}
                \label{fig:IMAGENET-VGG19-BN-FGSM}
        \end{subfigure}        
        \vspace{0.2em}
        \centering \crule[red!50!white!100]{1em}{0.5em} \footnotesize{(Orange):} $\mu = \{0.1, 0.00156\}$ 	  				\crule[purple!60!white!100]{1em}{0.5em} (Pink):  $\mu = \{0.15, 0.00195\}$	
        \crule[blue!50!pink!50!white!100]{1em}{0.5em} (Purple):  $\mu = \{0.2, 0.00293\}$								\crule[blue!50!white!100]{1em}{0.5em} (Blue): $\mu = \{0.05, 0.00117\}$
        
		\caption{Noise Effectiveness charts for FGSM attacks. Area under the blue color denotes NSS scores for the 			correctly classified samples for the given noise level. Orange, pink, and purple colors denoted the NSS 			scores of the samples that were successfully attacked. Each color denotes the noise level added to the 				dataset with respect to the corresponding attack, where in $\mu = \{\mu_1, \mu_2\}$, 				$\mu_1$ is the noise level added to MNSIT and $\mu_2$ is the noise level added to rest of the 			datasets.}
\label{fig:GSAeffect}
\end{figure*}

\begin{figure*}[ht]
        \begin{subfigure}[b]{0.33\textwidth}
                \includegraphics[width=\linewidth, height=30mm]{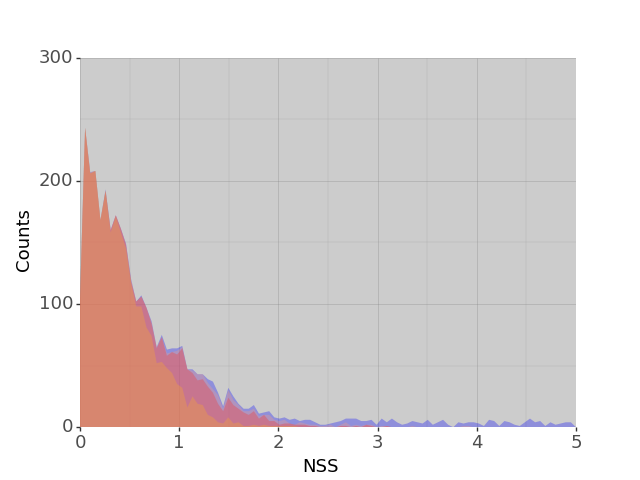}
                \caption{MNIST-MLP1-FGM}
                \label{fig:MNIST-MLP1-FGM}
        \end{subfigure}%
        \begin{subfigure}[b]{0.33\textwidth}
                \includegraphics[width=\linewidth, height=30mm]{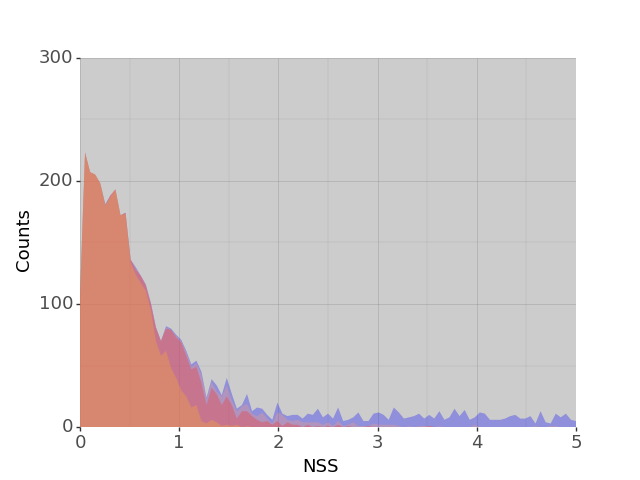}
                \caption{MNIST-MLP2-FGM}
                \label{fig:MNIST-MLP2-FGM}
        \end{subfigure}%
        \begin{subfigure}[b]{0.33\textwidth}
                \includegraphics[width=\linewidth, height=30mm]{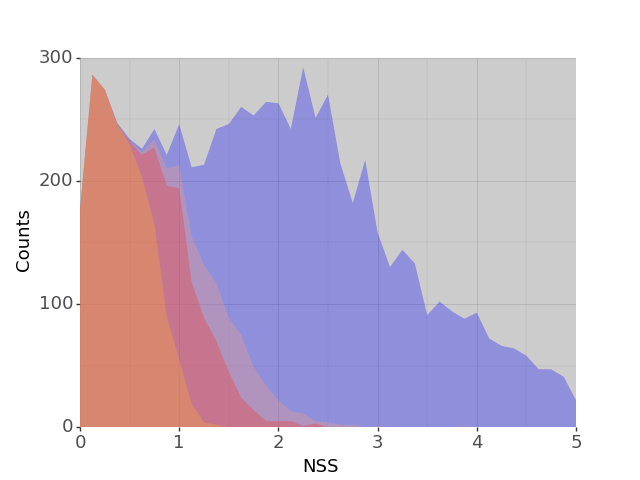}
                \caption{MNIST-LeNet-FGM}
                \label{fig:MNIST-LENET-FGM}
        \end{subfigure}\vspace{-0.15em}
        \begin{subfigure}[b]{0.33\textwidth}
                \includegraphics[width=\linewidth, height=30mm]{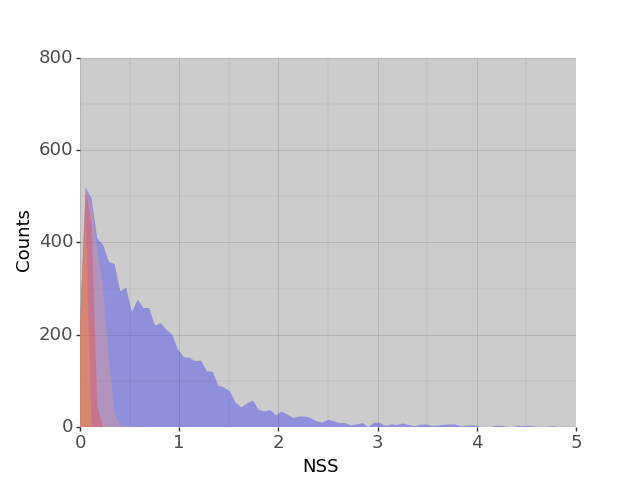}
                \caption{CIFAR10-DenseNet-40 w/o Bottleneck}
                \label{fig:CIFAR10-DenseNet-FGM}
        \end{subfigure}%
        \begin{subfigure}[b]{0.33\textwidth}
                \includegraphics[width=\linewidth, height=30mm]{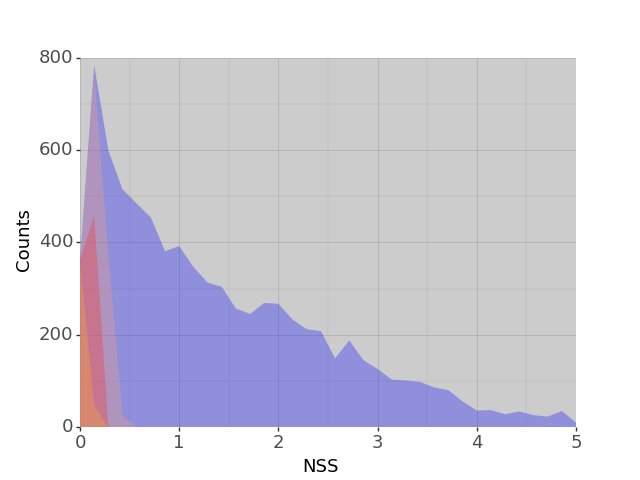}
                \caption{CIFAR10-ResNet18-FGM}
                \label{fig:CIFAR10-Res18-FGM}
        \end{subfigure}%
        \begin{subfigure}[b]{0.33\textwidth}
                \includegraphics[width=\linewidth, height=30mm]{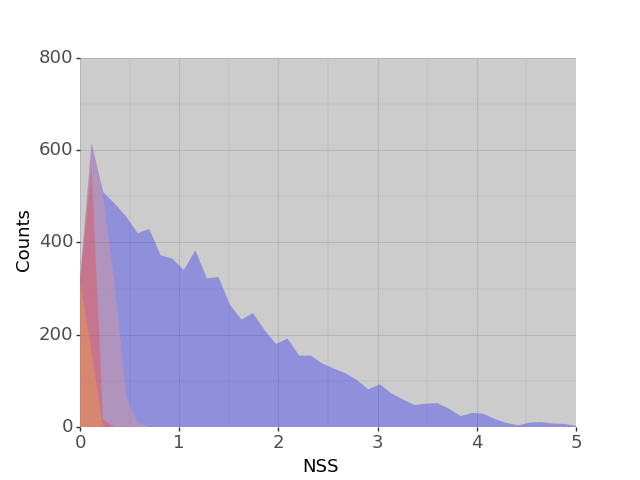}
                \caption{CIFAR10-VGG13-FGM}
                \label{fig:CIFAR10-VGG13-FGM}
        \end{subfigure}
        \begin{subfigure}[b]{0.33\textwidth}
                \includegraphics[width=\linewidth, height=30mm]{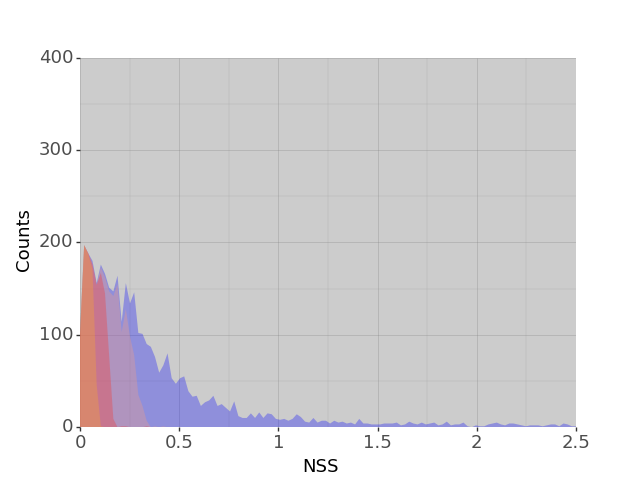}
                \caption{CIFAR100-DenseNet-40 w/o Bottleneck}
                \label{fig:CIFAR100-DenseNet-FGM}
        \end{subfigure}%
        \begin{subfigure}[b]{0.33\textwidth}
                \includegraphics[width=\linewidth, height=30mm]{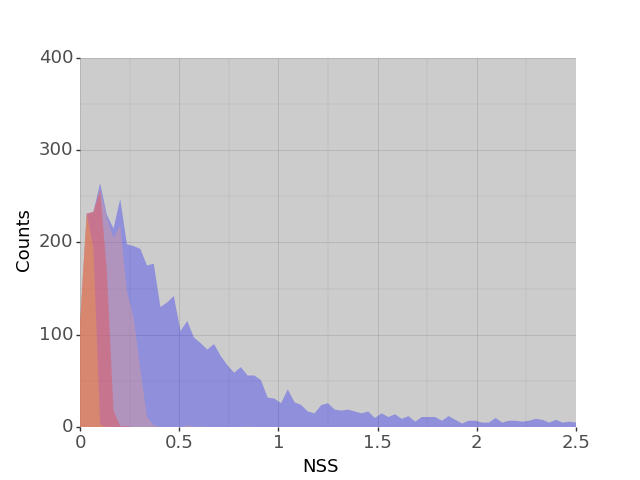}
                \caption{CIFAR100-ResNet18-FGM}
                \label{fig:CIFAR100-Res18-FGM}
        \end{subfigure}%
        \begin{subfigure}[b]{0.33\textwidth}
                \includegraphics[width=\linewidth, height=30mm]{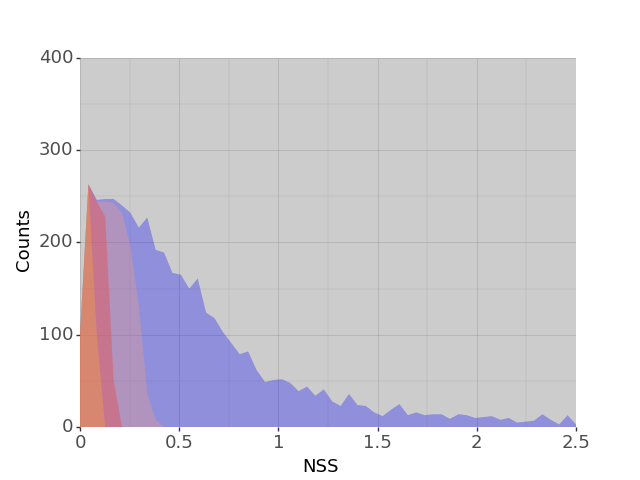}
                \caption{CIFAR100-VGG13 }
                \label{fig:CIFAR100-VGG13-FGM}
        \end{subfigure}
        \begin{subfigure}[b]{0.25\textwidth}
                \includegraphics[width=\linewidth, height=30mm]{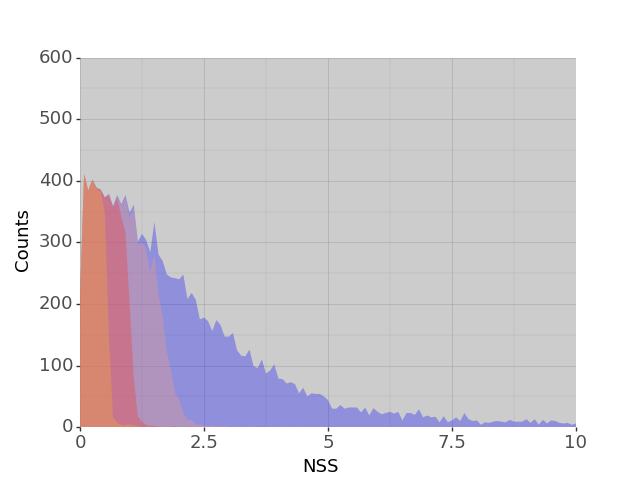}
                \caption{IMAGENET-DenseNet-161}
                \label{fig:IMAGENET-DENSE-161-FGM}
        \end{subfigure}%
        \begin{subfigure}[b]{0.25\textwidth}
                \includegraphics[width=\linewidth, height=30mm]{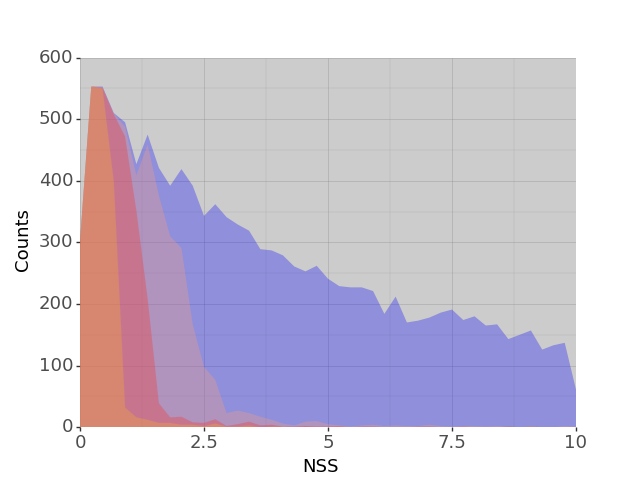}
                \caption{IMAGENET-Inception-V3}
                \label{fig:IMAGENET-INCEPTION-FGM}
        \end{subfigure}%
        \begin{subfigure}[b]{0.25\textwidth}
                \includegraphics[width=\linewidth, height=30mm]{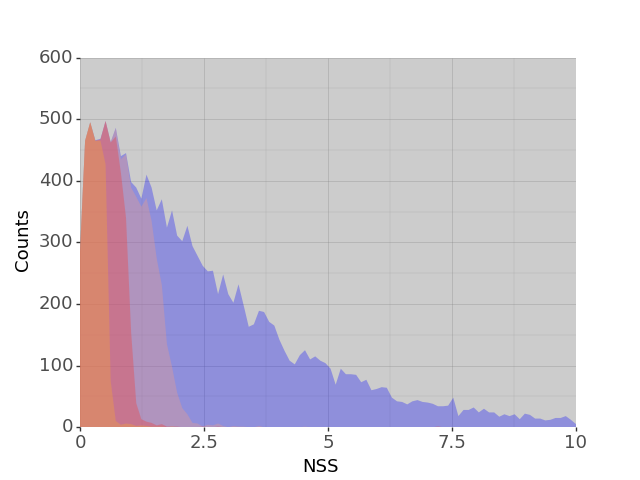}
                \caption{IMAGENET-ResNet-152}
                \label{fig:IMAGENET-RESNET-152-FGM}
        \end{subfigure}%
        \begin{subfigure}[b]{0.25\textwidth}
        \centering
                \includegraphics[width=\linewidth, height=30mm]{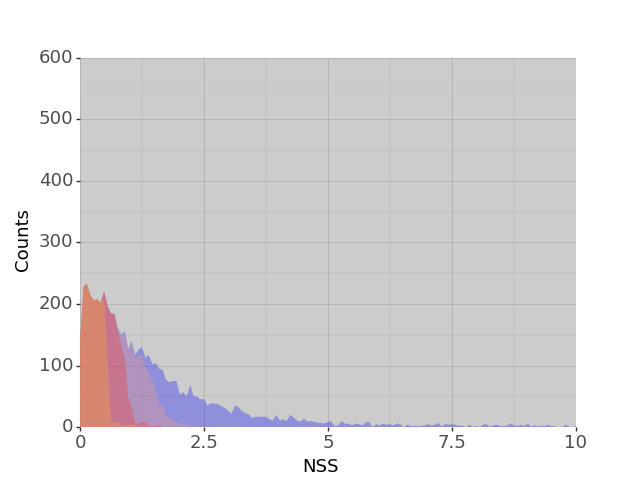}
                \caption{IMAGENET-VGG-19-BN}
                \label{fig:IMAGENET-VGG19-BN-FGM}
        \end{subfigure}
        \vspace{0.2em}
        \centering \crule[red!50!white!100]{1em}{0.5em} \footnotesize{(Orange):} $\mu = \{0.1, 0.00156\}$ 	  		\crule[purple!60!white!100]{1em}{0.5em} (Pink):  $\mu = \{0.15, 0.00195\}$	
        \crule[blue!50!pink!50!white!100]{1em}{0.5em} (Purple):  $\mu = \{0.2, 0.00293\}$								\crule[blue!50!white!100]{1em}{0.5em} (Blue): $\mu = \{0.05, 0.00117\}$
        
		\caption{Noise Effectiveness charts for FGM attacks. Area under the blue color denotes NSS scores for the 			correctly classified samples for the given noise level. Orange, pink, and purple colors denoted the NSS 			scores of the samples that were successfully attacked. Each color denotes the noise level added to the 				dataset with respect to the corresponding attack, where in $\mu = \{\mu_1, \mu_2\}$, 				$\mu_1$ is the noise level added to MNSIT and $\mu_2$ is the noise level added to rest of the 			datasets.}
		\label{fig:GAeffect}
\end{figure*}

\begin{figure*}[ht]
        \begin{subfigure}[b]{0.25\textwidth}
                \includegraphics[width=\linewidth, height=30mm]{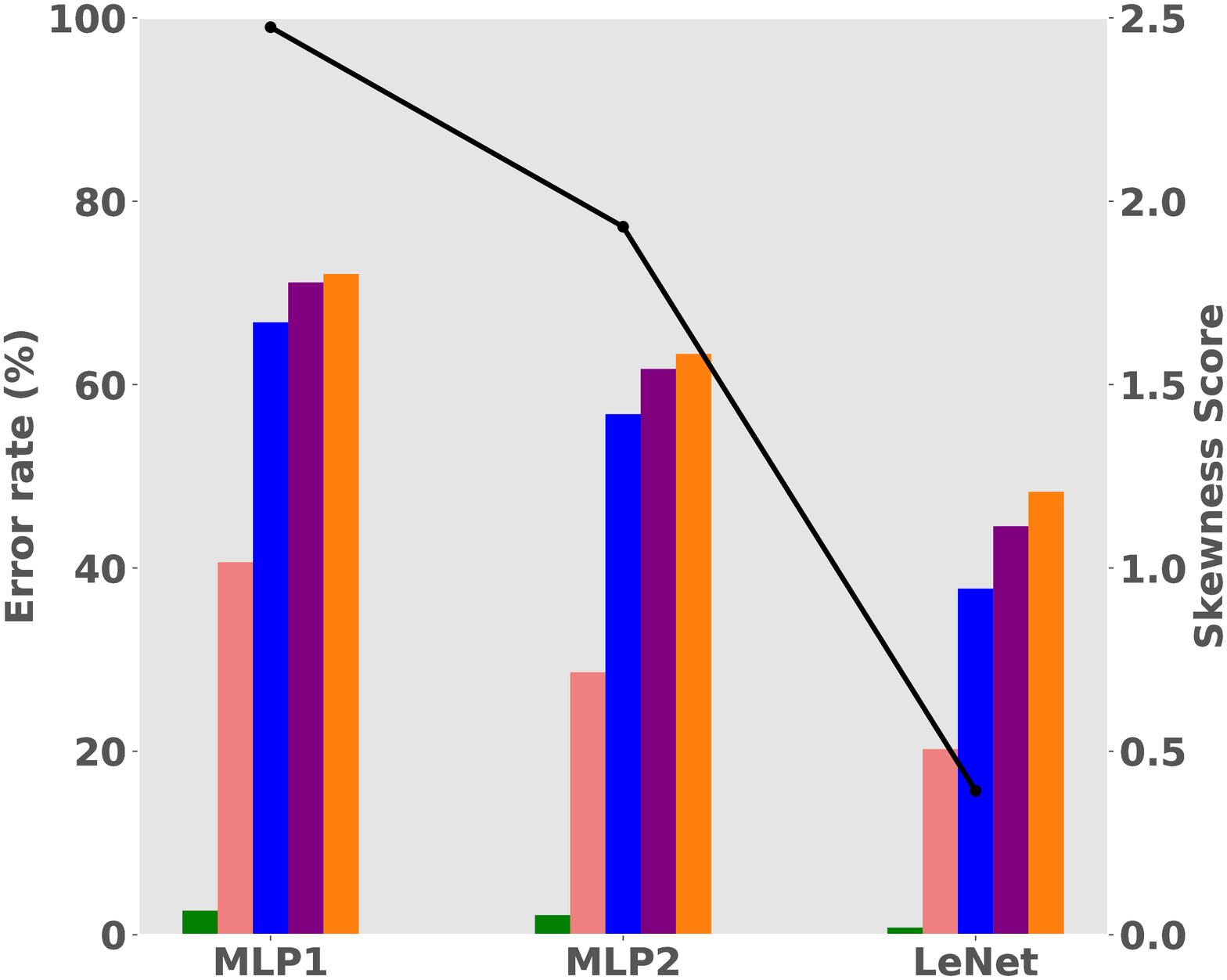}
                \caption{MNIST-FGM}
                \label{fig:mnist-ga}
        \end{subfigure}%
        \begin{subfigure}[b]{0.25\textwidth}
                \includegraphics[width=\linewidth, height=30mm]{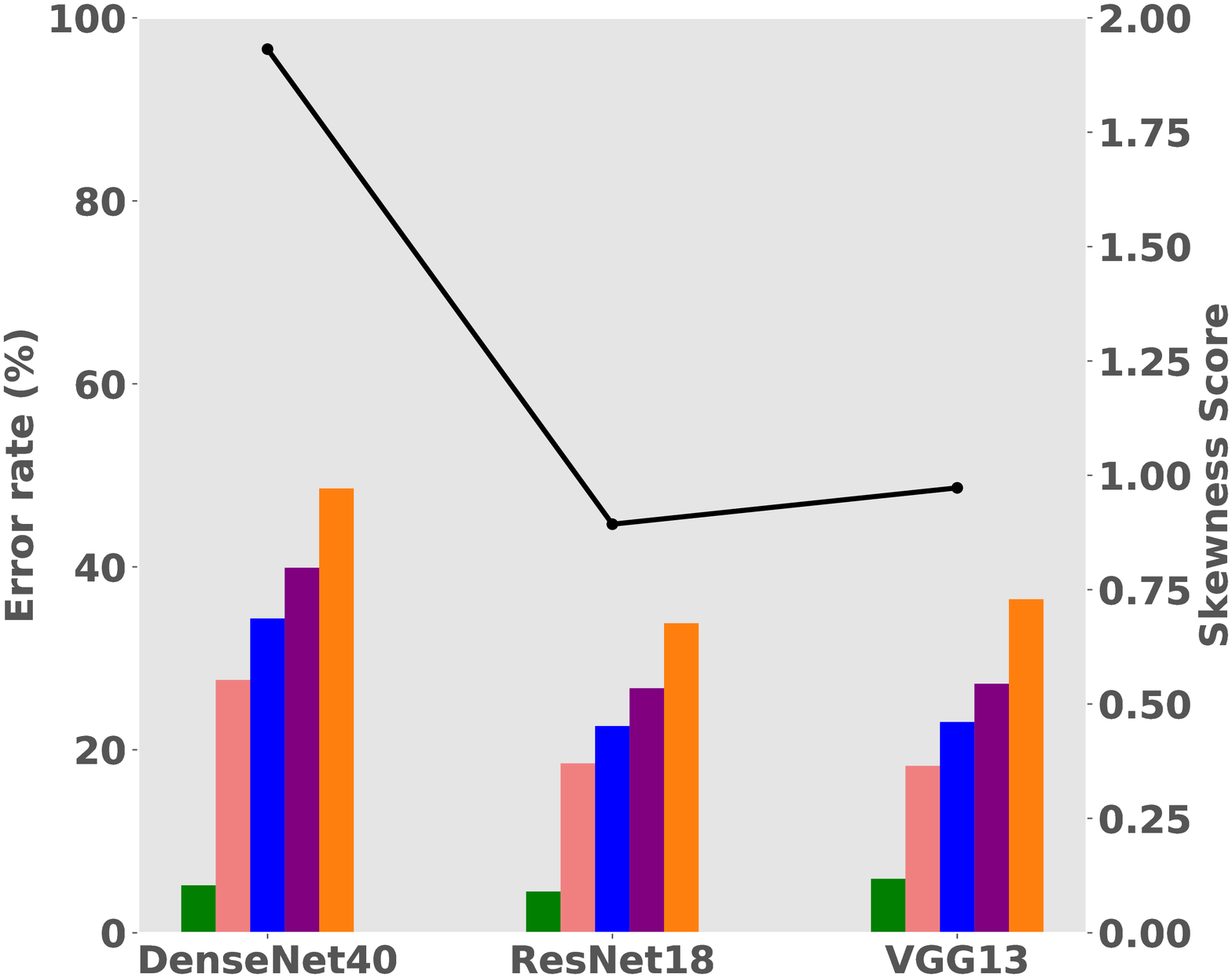}
                \caption{CIFAR10-FGM}
                \label{fig:cifar10-ga}
        \end{subfigure}%
        \begin{subfigure}[b]{0.25\textwidth}
                \includegraphics[width=\linewidth, height=30mm]{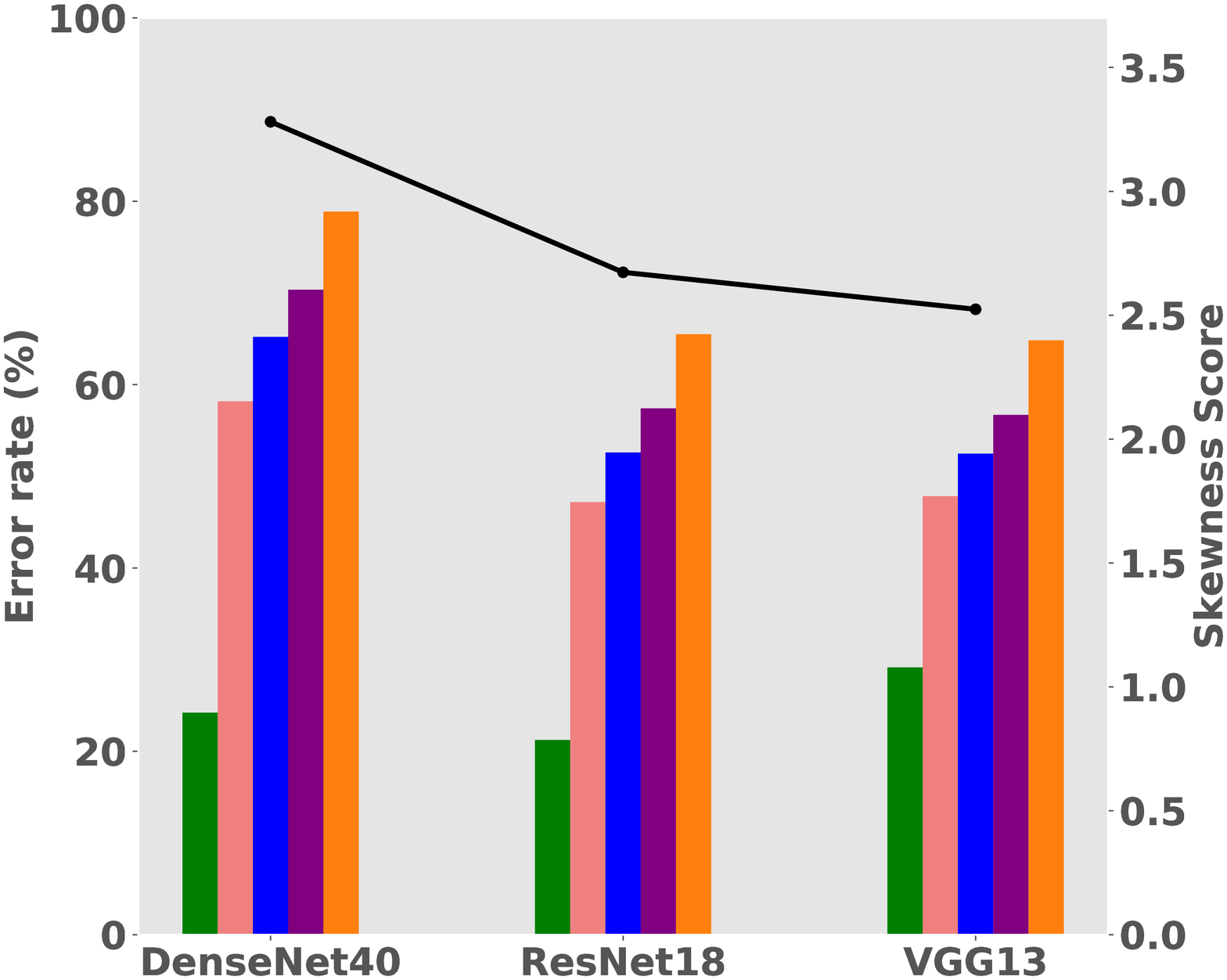}
                \caption{CIFAR100-FGM}
                \label{fig:cifar100-ga}
        \end{subfigure}%
        \begin{subfigure}[b]{0.25\textwidth}
                \includegraphics[width=\linewidth, height=30mm]{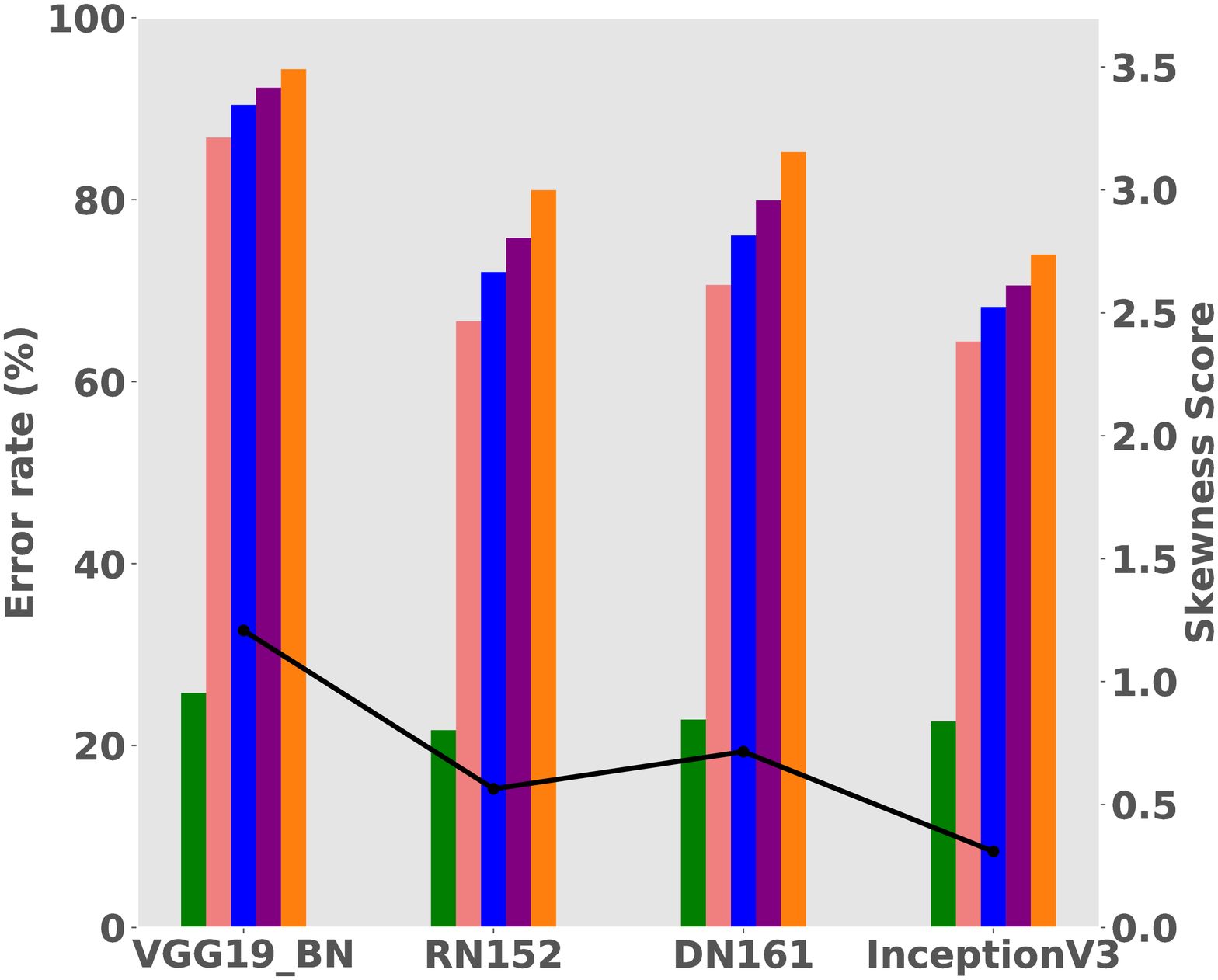}
                \caption{IMAGENET-FGM}
                \label{fig:imagenet-ga}
        \end{subfigure}
        \begin{subfigure}[b]{0.25\textwidth}
                \includegraphics[width=\linewidth, height=30mm]{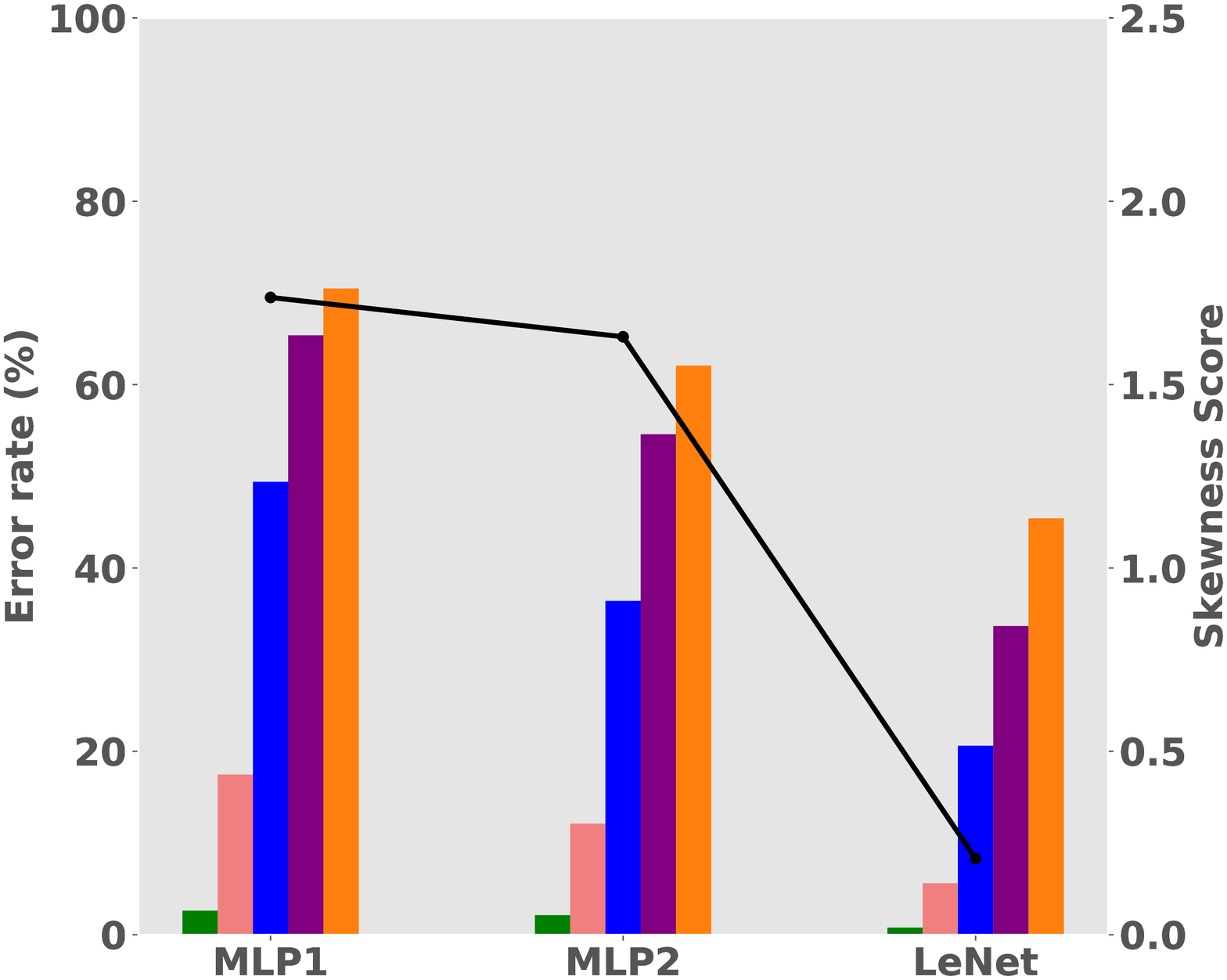}
                \caption{MNIST-FGSM}
                \label{fig:mnist-gsa}
        \end{subfigure}%
        \begin{subfigure}[b]{0.25\textwidth}
                \includegraphics[width=\linewidth, height=30mm]{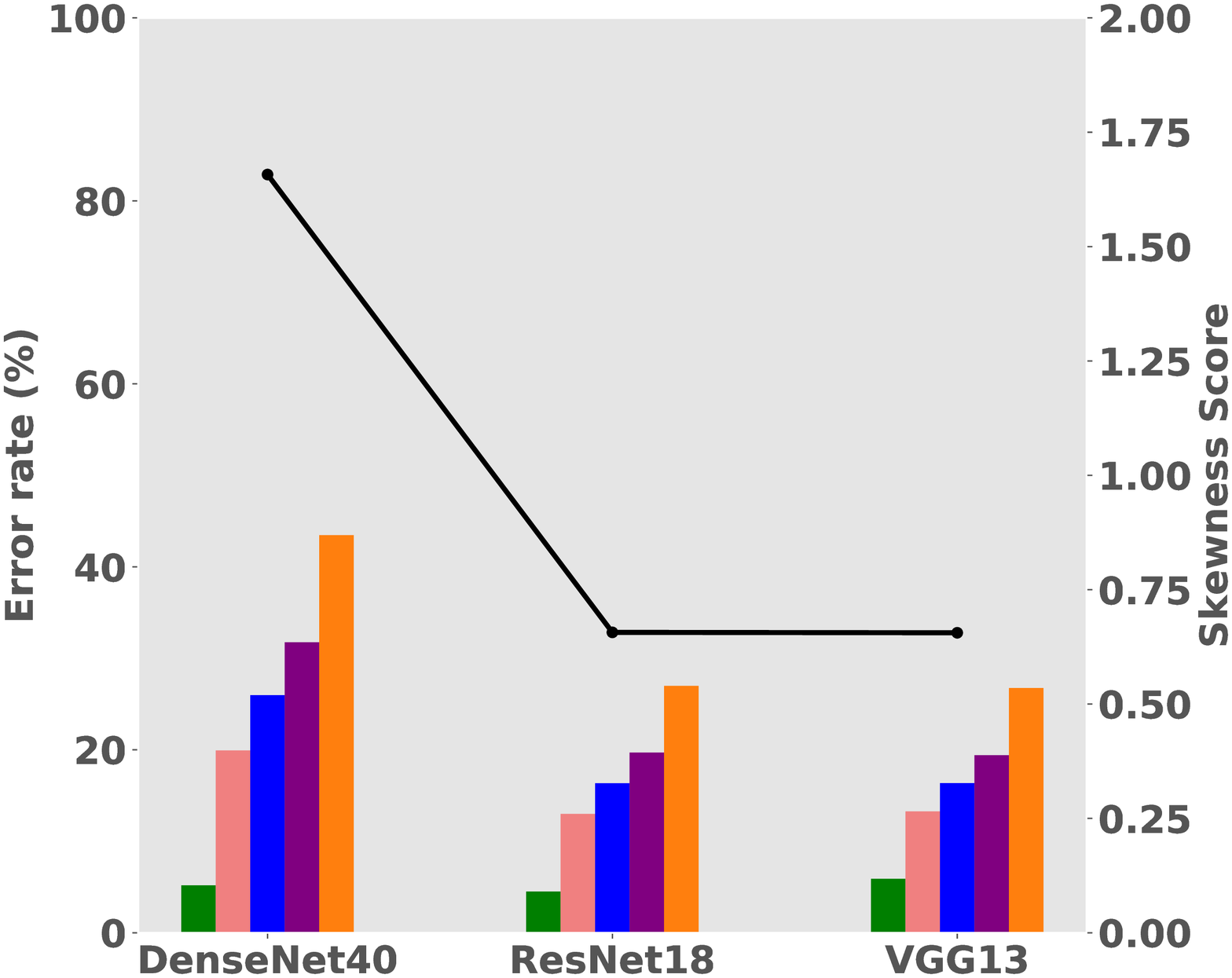}
                \caption{CIFAR10-FGSM}
                \label{fig:cifar10-gsa}
        \end{subfigure}%
        \begin{subfigure}[b]{0.25\textwidth}
                \includegraphics[width=\linewidth, height=30mm]{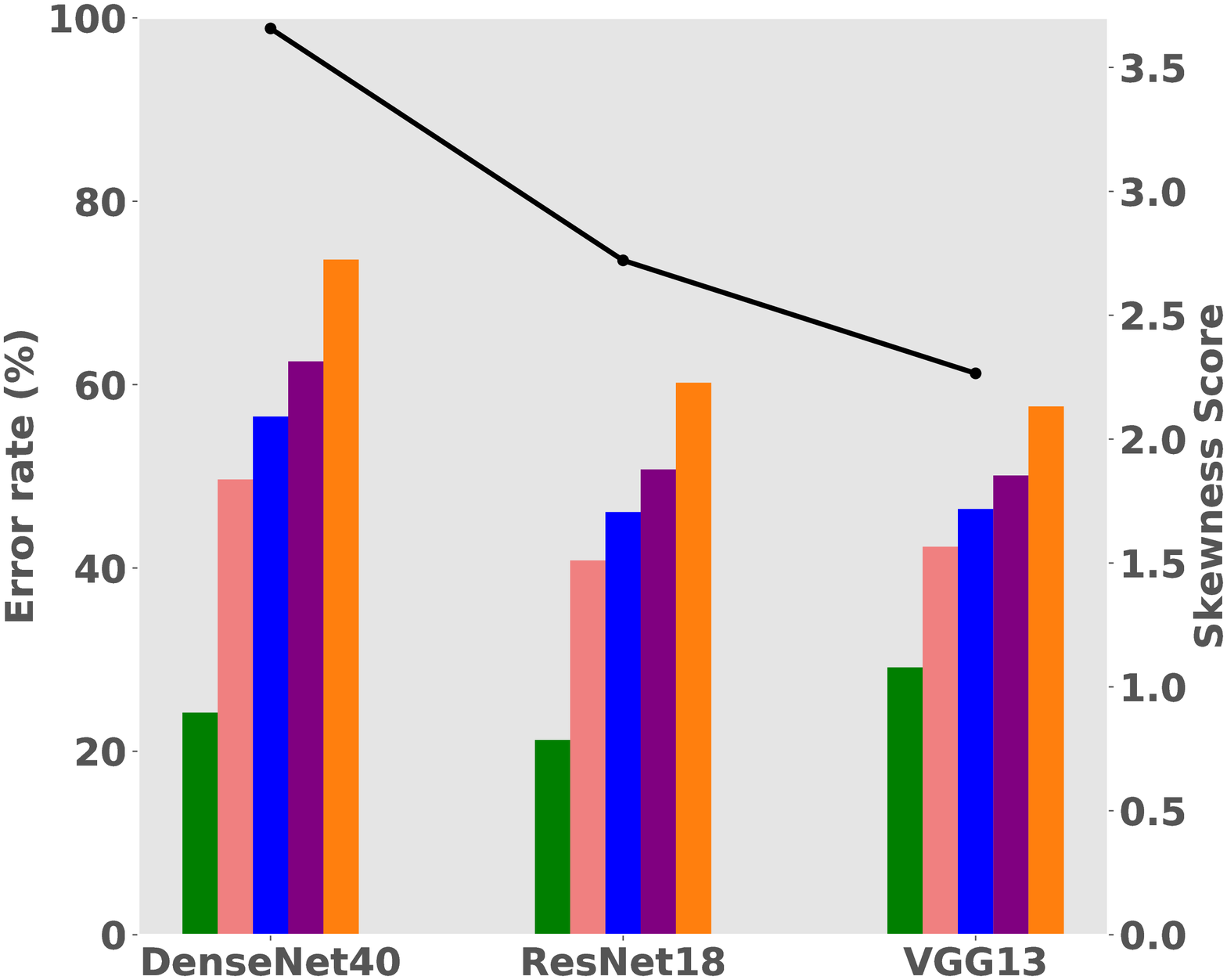}
                \caption{CIFAR100-FGSM}
                \label{fig:cifar100-gsa}
        \end{subfigure}%
        \begin{subfigure}[b]{0.25\textwidth}
                \includegraphics[width=\linewidth, height=30mm]{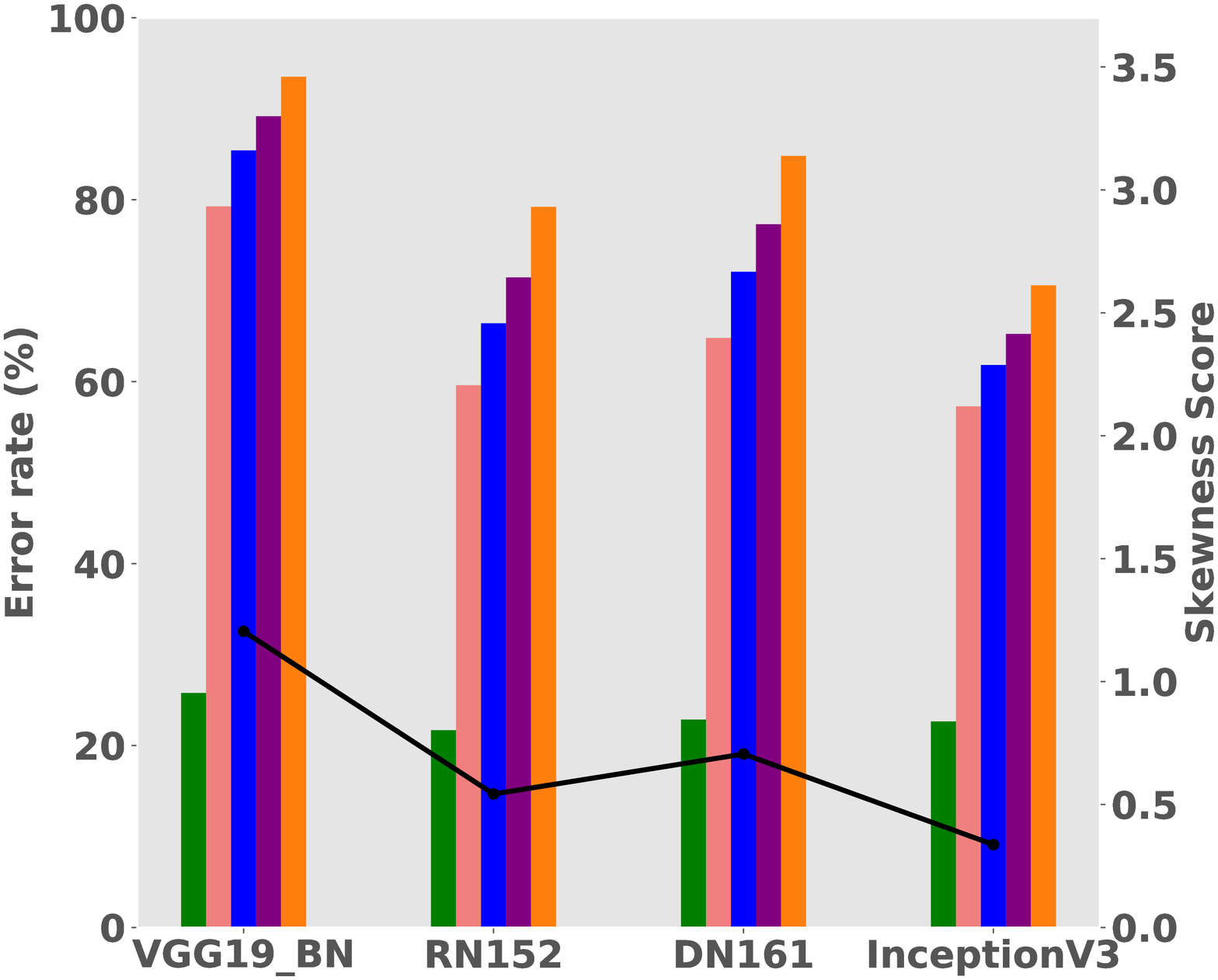}
                \caption{IMAGENET-FGSM}
                \label{fig:imagenet-gsa}
        \end{subfigure} 
        \vspace{0.25em}
		\centering        
        \crule[green!50!black!85]{1em}{0.5em} : \footnotesize{$\mu = \{0.0, 0.0\}$} \hspace{0.25em}\crule[purple!40!white!100]		{1em}{0.5em} : $\mu = \{0.05, 0.00117\}$ \hspace{0.25em} \crule[blue!100]{1em}{0.5em} : $\mu =			\{0.1, 0.00156\}$ \hspace{0.25em} \crule[violet!100]{1em}{0.5em} : $\epsilon = \{0.15, 0.00195\}$ 					\hspace{0.25em} \crule[orange!100]{1em}{0.5em} : $\mu = \{0.2, 0.00293\}$ 										\includegraphics[height=3mm]{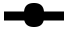} : Skewness score
        \vspace{0.5em}
        \caption{Skewness analysis for different architectures using both (a-d) Gradient Attacks and (e-h) Gradient Sign Attacks using a skewness threshold of $\tau = 5$. Each color denotes the noise level added to the dataset with respect to the corresponding attack, where in $\mu = \{\mu_1, \mu_2\}$,  $\mu_1$ is the noise level added to MNSIT and $\mu_2$ is the noise level added to rest of the datasets. The color bars represents the error rate of the respective architecture at the given noise and the circled line denotes the skewness score of the respective architecture.}
\label{fig:skewness}
\end{figure*}

\noindent $(3\times H \times W)$, where $H$ and $W$ are expected to be at least 224. The images are loaded in the range of $[0, 1]$ and then normalized using a $mean = [0.485, 0.456, 0.406]$ and $ std = [0.229, 0.224, 0.225]$ \cite{paszke2017automatic}.

\subsection{Validation Results}
\label{sec:vali_results}
It is important to substantiate both the effective and generalizable properties of the proposed metrics. The validation experiments for explaining these properties were designed accordingly.\vspace{0.25em}

\subsubsection{\textbf{Noise Sensitivity Score}}
\label{sssec: prop}
As explained in section \ref{sssec:DSM}, input samples with smaller NSS are more prone to be attacked successfully with an increase in the noise level $\epsilon$. Given the maximum allowable noise level $\mu$, the adversarial algorithm finds the smallest $\epsilon$ which perturbs the original sample in order to mis-classify it. In order to validate correctness of the exception, we design the following experiment. Firstly, we apply FGSM/FGM to a dataset with an initial noise level, which is set $0.05$ for MNIST and $0.00117\ (0.3/256)$ for other three datasets. The estimated NSSs of all the correctly classified inputs are recorded and indicated as blue color in the Fig.~\ref{fig:GSAeffect} and Fig.~\ref{fig:GAeffect}. It is worth to mention that the initial noise level can be of any value larger than or equal to $0$, and the initial value does not  impact the proposed metrics. The main reason of not setting the noise level to zero is because we want to show a DNN's adversarial accuracy, obtained from the low noise level, is not an ideal adversarial robustness metric. Besides less explainable, it does not always accurately predict the performance of the DNN in the context of a higher allowable noise level. Secondly, in order to show the consistent performance of the proposed NSS, we increase the allowable noise level to three different levels and record the NSSs for those who are successfully attacked (i.e., misclassified). The three different levels are $0.1$, $0.15$, $0.2$ for MNIST and $0.00156 (0.4/256)$, $0.00195 (0.5/256)$, $0.00293 (0.75/256)$ for other three datasets, and the corresponding NSSs are marked in orange, pink and purple Fig.~\ref{fig:GSAeffect} and Fig.~\ref{fig:GAeffect} respectively. The obtained NSSs from different DNN architectures across different datasets under the FGSM and FGM are presented as histogram charts and shown in Fig.~\ref{fig:GSAeffect} and Fig.~\ref{fig:GAeffect} respectively. As we can see, for all the datasets and architectures, the overlapping regions are always on the very left side of the respective charts. In other words, all the charts show the inputs with lower NSS are easier to be attacked successfully. One can expect the mis-classification to increase with a higher allowable noise level. This can also be observed from all the figures in Fig.~\ref{fig:GSAeffect} and Fig.~\ref{fig:GAeffect}, where the overlapping regions are getting larger with an increase in the allowable noise level. In addition, the above observations are consistent for both FGSM and FGM. These results are all strong evidences for the effectiveness and generalization of the proposed NSS. More interestingly, based on the distribution of the obtained NSSs, we are able to gain a relatively precise intuition about different DNNs' performance on a dataset. For example, based on the Fig.~\ref{fig:CIFAR10-DenseNet-FGSM} and Fig.~\ref{fig:CIFAR10-Res18-FGSM}, it is obvious that ResNet-18 gives a better adversarial performance than DenseNet-40-without-bottleneck since the recorded NSSs of ResNet-18 spread in a much wider range than the ones of DenseNet-40-without-bottleneck. Following the same theory, based on the Fig.~\ref{fig:CIFAR10-Res18-FGSM} and Fig.~\ref{fig:CIFAR100-Res18-FGSM}, we can predict the ResNet-18 is more robust on CIFAR-10 than CIFAR-100. These two observations can be confirmed by the corresponding adversarial error rates shown in Table~\ref{tab:skewness}. The boundary between different color regions are also worth to pay attention. As we can see, except for MLP1 and MLP2, the boundaries in all other charts of Fig.~\ref{fig:GSAeffect} and Fig.~\ref{fig:GAeffect} are all relatively vertical. However, we can see the boundaries between purple and blue regions has shallower steepness than other two boundaries in general, which implies the NSS works more effectively with a lower allowable noise level. It confirms the $DJM$'s limitation as claimed in the section~\ref{sssec:cf}. \vspace{0.25em}

\subsubsection{\textbf{NSS Skewness based Adversarial Robustness Metric}}
\label{sssec: val_NSS}
We validate the proposed skewness based dataset robustness metric by comparing the adversarial error rate for all the architectures mentioned in section \ref{sec:dataArch} under both FGSM and FGM attacks. The same experimental setup of validating NSS is used. In Table \ref{tab:skewness},  we present an extensive set of results of all the architectures mentioned in \ref{sec:dataArch} from all datasets. It is to be noted that although the validation performance of these architectures are very close to each other, the adversarial performances varies a lot. This observation puts forward an interesting question as to how differently an architecture learns a dataset? Also, can adversarial accuracy be used as the metric to evaluate that? We argue that the adversarial error rate is not the ideal candidate for measuring robustness. 

One interesting observation across all the results in the table is that the skewness score is directly proportional to the adversarial error rate. It is to be noted that the skewness scores was calculated for the lowest noise level for each of the dataset, but the values correlate with the architecture's performance at higher level noises. This characteristics is not true for adversarial accuracies at lowest noise level. For instance, in the CIFAR10-FGM case from Table \ref{tab:skewness}, VGG13 architecture showed lower error rate at lower noise level as compared to ResNet-18 but the trend flipped when we increased the level of noise. Similar observations were witnessed in CIFAR-100 FGM and FGSM results. These results substantiate the prediction property of the skewness score. Additionally, we can clearly see the correlation of the skewness scores with the error rates for across all architectures and datasets in Figure \ref{fig:skewness}. 

It is seen in Table \ref{tab:skewness}, that the skewness scores are not biased to a certain dataset. It retains its unique property across all the datasets. The skewness score is highest for the architecture having highest error rate. Furthermore, it is surprising to observe how easy it is to mislead a network trained on large datasets like ImageNet. Another important observation, is to note the difference in adversarial performance using FGM and FGSM attacks. This difference is prominently seen from the adversarial accuracies across all datasets. In general, we notice that FGM is a more powerful attack when compared to FGSM.
\begin{table}[H]
\captionsetup{font=footnotesize}
\caption{In this table, we show the Validation (VA) error rate and Adversarial (AD) error rate from different setups. The corresponding skewness based dataset robustness scores are also shown in the last column.}
\resizebox{\columnwidth}{!}{%
\scalebox{0.8}{%
\begin{tabular}{ccccccccc}
\hline
Dataset & \makecell{Attack \\ type} & Models & VA \% & \makecell{AD \% \\ $\mu=0.05$} & \makecell{AD \%\\ $\mu=0.1$} & 
\makecell{AD \%\\ $\mu=0.15$} & \makecell{AD \% \\ $\mu=0.2$} & \makecell{Skewness \\ score}\\
\hline
\multirow{8}{*}{MNIST} & \multirow{3}{*}{FGM} & MLP1 & {2.61\%} & {40.63\%} & {66.79\%} & {71.14\%} & {72.06\%} & {2.4747}\\
& & MLP2 & {2.13\%} & {28.62\%} & {56.78\%} & {61.71\%} & {63.35\%} & {1.9306}\\ 
& & LeNet & {0.77\%} & {20.24\%} & {37.74\%} & {44.55\%} & {48.30\%} & {0.3923}\\ 
&  &  &  &  &  &  &  &\\
& \multirow{3}{*}{FGSM} & MLP1 & {2.61\%} & {17.46\%} & {49.40\%} & {65.38\%} & {70.48\%} & {1.7377}\\
& & MLP2 & {2.13\%} & {12.11\%} & {36.41\%} & {54.58\%} & {62.07\%} & {1.6304}\\ 
& & LeNet & {0.77\%} & {5.62\%} & {20.60\%} & {33.66\%} & {45.41\%} & {0.2072}\\ [0.5ex]
\hline
\multirow{8}{*}{CIFAR-10} & \multirow{3}{*}{FGM} & DenseNet40 & {5.18\%} & {27.64\%} & {34.35\%} & {39.90\%} & {48.57\%} & {1.9316}\\ [0.5ex]
& & ResNet18 & {4.51\%} & {18.52\%} & {22.60\%} & {26.72\%} & {33.83\%} & {0.8931}\\
& & VGG13 & {5.90\%} & {18.24\%} & {23.03\%} & {27.22\%} & {36.45\%} & {0.9726}\\ [0.5ex]
&  &  &  &  &  &  &  &\\
& \multirow{3}{*}{FGSM} & DenseNet40 & {5.18\%} & {19.93\%} & {25.97\%} & {31.75\%} & {43.46\%} & {1.6575}\\ [0.5ex]
& & ResNet18 & {4.51\%} & {12.99\%} & {16.35\%} & {19.70\%} & {26.98\%} & {0.6567}\\
& & VGG13 & {5.90\%} & {13.26\%} & {16.36\%} & {19.41\%} & {26.75\%} & {0.6558}\\ [0.5ex]
\hline
\multirow{8}{*}{CIFAR-100} & \multirow{3}{*}{FGM} & DenseNet40 & {24.22\%} & {58.18\%} & {65.21\%} & {70.35\%} & {78.88\%} & {3.2808}\\ [0.5ex]
& & ResNet18 & {21.24\%} & {47.19\%} & {52.59\%} & {57.41\%} & {65.49\%} & {2.6733}\\
& & VGG13 & {29.15\%} & {47.84\%} & {52.47\%} & {56.70\%} & {64.83\%} & {2.5241}\\ [0.5ex]
& &  &  &  &  &  &  &\\
& \multirow{3}{*}{FGSM} & DenseNet40 & {24.22\%} & {49.66\%} & {56.52\%} & {62.53\%} & {73.64\%} & {3.6574}\\
& & ResNet18 & {21.24\%} & {40.82\%} & {46.09\%} & {50.75\%} & {60.20\%} & {2.7215}\\
& & VGG13 & {29.15\%} & {42.32\%} & {46.44\%} & {50.08\%} & {57.62\%} & {2.2655}\\ 
\hline
\multirow{8}{*}{IMAGENET} & \multirow{3}{*}{FGM} & VGG19\_bn & {25.78\%} & {86.83\%} & {90.42\%} & {92.31\%} & {94.35\%} & {1.2080}\\
& & ResNet152 & {21.69\%} & {66.62\%} & {72.05\%} & {75.81\%} & {81.04\%} & {0.5638}\\ 
& & DenseNet161 & {22.86\%} & {70.63\%} & {76.07\%} & {79.92\%} & {85.23\%} & {0.7149}\\
& & InceptionV3 & {22.65\%} & {64.40\%} & {68.21\%} & {70.57\%} & {73.95\%} & {0.3091}\\ 
 & &  &  &  &  &  &  &\\
& \multirow{3}{*}{FGSM} & VGG19\_bn & {25.78\%} & {79.26\%} & {85.42\%} & {89.18\%} & {93.52\%} & {1.2047}\\
& & ResNet152 & {21.69\%} & {59.61\%} & {66.41\%} & {71.45\%} & {79.21\%} & {0.5428}\\
& & DenseNet161 & {22.86\%} & {64.80\%} & {72.08\%} & {77.30\%} & {84.81\%} & {0.7055}\\
& & InceptionV3 & {22.65\%} & {57.28\%} & {61.83\%} & {65.25\%} & {70.58\%} & {0.3371}\\
\hline
\end{tabular} 
}
}
\vspace{-1em}
\label{tab:skewness}
\end{table}

\section{Discussion}
\label{sec:discussion}
We propose a new class of algorithms that can quantize noise sensitivity of an input under a DNN fix-directional adversarial attack. The proposed NSS and skewness based dataset adversarial robustness metric are extensively tested using popular DNN architectures and the results approve that they work effectively and consistently across all the four state-of-the-art datasets. Compared to previous works, our validation experiments are much more thorough and convincing. Furthermore, an important characteristic of most of the previously proposed robustness metrics is that they all work in the input domain. This basically comes down to taking some form of distance metrics in terms of $L_{0}, L_2$ or $L_{\infty}$  norm and finding the distance between the original and the adversarial images. We propose to measure the robustness score of a given framework by getting the scores directly from the adversarial architecture attack mathematical model.  Although we focused our work on the misclassification adversarial attack, the proposed metrics are also applicable to targeted misclassification attacks. To do that, we only need to change $y_t$ in Eq.~\eqref{eq:s_score} from predicted class error to the targeted class error. In addition, the proposed metrics are also suitable to non-gradient based attack, such as Gaussian noise attack and uniform noise attack \cite{rauber2017foolbox}.

The proposed metrics provide an insightful mathematical point of view to look at a DNN's dataset-wise robustness under a fix-directional adversarial attack and quantizes the predicted robustness. One of our key results is the absence of an absolute robustness concept for a DNN since applying the same DNN to different datasets may lead to significant robustness gaps. Therefore, we argue that understanding a specific sample's robustness is more important than measuring a DNN's robustness to a specific dataset. The proposed NSS is designed for this purpose. Interestingly, even without the proposed skewness based dataset robustness metric, we can already have an intuition about a DNN's adversarial performance on a dataset based on the plotted distribution of the NSSs, such as the charts in Fig~\ref{fig:GSAeffect} and Fig~\ref{fig:GAeffect}. One limitation of NSS is it may not accurately reflect the noise sensitivity of a DNN's input with a large allowable maximum noise level. The larger allowable maximum noise level results in a larger neighborhood on an embedded surface $m_i$ where the $DJM$ may not be able to provide a desired accuracy level approximation. In other words, the nonlinearity of the enlarged neighborhood may not be approximated accurately with a single piecewise linear approximation. If that is the case, a multi-step NSS approach can be used to solve the problem. Instead of using the $s_i$ defined in Eq.~\eqref{eq:DJM_speed}, we can divide the targeted large noise level into multiple small noise levels and apply Eq.~\eqref{eq:DJM_speed} for each of them. Then, the final rate of change is the average of the values estimated from those small noise levels. However, due to the perceptual constraint, the maximum allowable noise level cannot be significantly large, especially for the FGM. Hence, it is uncommon to use the proposed multi-step NSS. 

Our future works will focus on extending the NSS to make it work under an unfix-directional attack. Another interesting direction to look into is leveraging the proposed NSS during a training process in order to improve a DNN's adversarial performance.

\section{Related Work}
\label{sec:relwork}
Behavior of machine learning algorithms, precisely deep learning, in an adversarial environment is an active area of research in the security community. Various adversarial attacks have been devised by either directly using the respective model details or by transferring the generated adversarial samples from one model to another \cite{papernot2017practical}. Over the years, different techniques are developed for crafting adversarial samples using the gradient procedure \cite{goodfellow2014explaining,szegedy2013intriguing}. These attacks were discussed in detail in section \ref{related:1}. On the other hand, many attempts have been made at making DNN robust to these adversarial attacks. Recent studies have focused on detecting adversarial examples directly \cite{feinman2017detecting,metzen2017detecting,grosse2017statistical}. Moreover, these techniques also require modifying the model or acquiring sufficient adversarial examples, such as training new sub-models \cite{feinman2017detecting}, retraining a revised model as a detector using known adversarial examples \cite{metzen2017detecting} or performing a statistical test on a large group of adversarial and benign examples \cite{grosse2017statistical}. Some classical adversarial defenses are discussed in section \ref{related:2} but further details on different defense mechanisms falls outside the scope of this paper.

With the development of different adversarial crafting and defensive algorithms, there is a need to introduce a robustness metric that measures the resiliency of machine learning algorithms to adversarial perturbations. Many researchers consider the adversarial accuracy of an algorithm as the measure for evaluating robustness \cite{tabacof2016exploring}. Also, different distance based norm metrics using $L_{0}$, $L_{2}$ and $L_{\infty}$ are used for robustness analysis. A detailed analysis of these robustness metrics is provided in section \ref{related:3}. In this paper, we proposed a skewness based dataset robustness metric and validated the results in a cross-framework and cross-dataset scenarios. 

\section{Conclusion}
\label{sec:conclusion}
In this work, we proposed the Noise Sensitivity Score (NSS) which measures the sensitivity of different inputs to the same noise level. The NSS together with the skewness based dataset robustness metric aids in providing a resilient metric which works for different fix-directional gradient attacks across various architectures and datasets. We performed a comprehensive set of experiments to substantiate the effectiveness of our proposed metrics. 

\bibliographystyle{IEEEtran}
\bibliography{reference}

\end{document}